%% file: journal_final_16.tex
\documentclass[twocolumn,doublespacing ,10pt,journal]{IEEEtran}
 \normalsize
\usepackage{epsfig}
\usepackage{amsmath}
\usepackage{theorem}
\usepackage{amsmath}
\usepackage{mathrsfs}
\usepackage{epsfig}
\usepackage{amsfonts}
\usepackage{amssymb}
\usepackage{tabularx}
\usepackage{mathrsfs}
\usepackage{euscript}
\usepackage{graphicx}
\usepackage{multirow}
\usepackage{stfloats}
\usepackage{cite}
\usepackage{algorithm}
\usepackage{color}

\usepackage{subcaption}
\usepackage{subcaption}

\usepackage{subcaption}
\usepackage[T1]{fontenc}
\usepackage{booktabs}
\usepackage{pgfplots}
\usepackage{grffile}
\pgfplotsset{compat=newest}
\usetikzlibrary{plotmarks}
\usetikzlibrary{arrows.meta}
\usepgfplotslibrary{patchplots}
\usepackage{amsmath,color}

\DeclareMathOperator{\erfc}{erfc}

\begin{document}
	%\bstctlcite{IEEEexample:BSTcontrol}
	\title{Analysis and Rate Optimization of GFDM-based Cognitive Radios }
	\author{A. Mohammadian, M. Baghani, C. Tellambura,~\IEEEmembership{Fellow,~IEEE}}
    
  %  \author{\IEEEauthorblockN{Amirhossein Mohammadian}
%\IEEEauthorblockA{Department of Electrical \\and Computer Engineering\\
%University of Alberta, \\Edmonton, Alberta T6G 2V4, Canada\\
%Email: am11@ualberta.ca}
%\and
%\IEEEauthorblockN{Mina Baghani}
%\IEEEauthorblockA{Electrical Engineering Department\\
%Amirkabir University of Technology,\\ Tehran, Iran\\
%Email: baghani@aut.ac.ir}
%\and
%\IEEEauthorblockN{Chintha Tellambura, Fellow, IEEE}
%\IEEEauthorblockA{Department of Electrical \\and Computer Engineering\\
%University of Alberta, \\Edmonton, Alberta T6G 2V4, Canada\\
%Email: chintha@ece.ualberta.ca}}

%	\thanks{$^{(1)}$ Microwave / Millimeter-Wave and Wireless Communications Research Lab., Electrical Engineering Department, Amirkabir University of Technology, Tehran, Iran (e-mail: {amirmohammadian, baghani}@aut.ac.ir).}

%}
\markboth{}%
{}	
	\maketitle	
	\begin{abstract}
	
	Generalized frequency division multiplexing (GFDM)  is suitable for cognitive radio (CR)  networks due to its low out-of-band (OOB) emission and high spectral efficiency. In this paper, we thus  consider the use of GFDM to allow an unlicensed secondary user (SU) to access a spectrum hole. However, in  an extremely congested spectrum scenario, both active  incumbent primary users (PUs)  on the  left and right channels of the spectrum hole will  experience OOB interference. While constraining this interference, we  thus investigate the problem of  power allocation to the SU transmit  subcarriers in order to maximize the overall data  rate where the SU receiver is employing Matched filter (MF) and zero-forcing (ZF) structures.  The  power allocation problem  is thus  solved as a classic convex optimization problem. Finally, total transmission rate of GFDM is compared with that of orthogonal frequency division multiplexing (OFDM).  For instance, when right and left interference temperature should be below  10\;dBm, the capacity gain of GFDM over OFDM is 400$\%$. 
	\end{abstract}
	

	\IEEEpeerreviewmaketitle
	
	%\linespread{1}
	\section{Introduction}
	
	 \IEEEPARstart{T}{he} development of fifth generation (5G) wireless networks faces the challenge of  congested and limited  wireless spectral resources\cite{magh17}. The main reasons for that are the massive growth of wireless data traffic and the assignment of almost all  spectrum  bands  below  6-GHz bands to   existing  wireless and cellular  applications. However, primarily due to usage patterns,  many spectrum bands temporarily become spectrum  holes. A  spectrum hole is a  free frequency band in a certain location  in which the  licensed (primary) users are not transmitting  temporarily. Such spectrum holes can be accessed by unlicensed  users or secondary users (SUs) under the  interweave cognitive radio (CR) paradigm.

%\begin{figure}
%\centering
%\input{my.tex}
%\caption{ My first matlab2tikz figure }
%\label{fig:myfirstfig}
%\end{figure} 

In 5G, the physical layer (PHY) maybe based on orthogonal frequency division multiplexing (OFDM) or Generalized frequency devision multiplexing (GFDM)\cite{magh18}. Although OFDM is robust against  frequency selective fading, its high out-of-band (OOB) interference may render it unsuitable for CR networks  [3]-[6].  GFDM has thus been proposed   \cite{magh2,magh16}. GFDM uses   multiple symbols per subcarrier  and shapes  each subcarrier  by a  circularly-shifted  prototype filter.  The spectral efficiency of GFDM is higher than that of OFDM because  the former  uses only single cyclic prefix (CP) for an entire block. As well,  GFDM can reduce the latency of PHY layer \cite{magh3}, the  main requirement of Tactile Internet \cite{magh28}. In addition, due to the shaping of each subcarrier individually, the OOB emission of GFDM is low \cite{magh4} and can  be further reduced by better design of filtering  \cite{magh5}. Multicarrier signaling methods are especially vulnerable against the  carrier frequency offset (CFOs) between transmitter and receiver, which arise due to Doppler effects, thermal effects, aging and others. In OFDM, a  CFO kills the perfect orthogonality among all the subcarriers, resulting in the lowering of the signal to interference ratio (SIR).   In contrast, in GFDM, a receiver filter can improve robustness  against CFOs \cite{magh22}, which maximizes the SIR. Thus, these advantages ensure that GFDM is an  attractive modulation method for 5G and CR networks [14], \cite{magh24}. However, GFDM incurs additional implementation complexity which has been improved in \cite{magh6,magh7,magh20,magh21}.  Due to these  advantages, especially the low OOB emissions, we consider the use of GFDM for unlicensed CR users in this paper.
	 
Unlicensed SUs  may access  the  primary user (PU) spectrum in two different modes. First, in underlay mode, they access  simultaneously with active PU transmissions,  but  ensure that the resulting interference on PU nodes  is  less  than a  specified interference threshold. Thus, in this mode, dynamic interference management is the key -- which can be achieved by several techniques such as secondary transmit power control, guard regions and/or proactive interference cancellation. Unfortunately, these techniques will  constrain the achievable SU rates. Moreover,  the burden of implementing such techniques falls on the secondary network. Second, in the  interweave  mode, the  SUs access spectrum holes only  \cite{magh8}. And that is the scenario investigated  in this paper. The challenge, however, is the accurate and dynamic sensing of  spectrum holes. Two common  sensing methods are energy detection and cyclostationary detection. \cite{magh9} reveals GFDM has a better complementary receiver operating characteristic (ROC)  compared to  OFDM based on energy detection method. In addition, Reference  \cite{magh10} shows signal detection improves with  GFDM due to its cyclostationary autocorrelation properties compared to OFDM. In light of these advantages, GFDM appears as a suitable candidate for interweave CR networks.

Although optimal power allocation  improves the SU  network performance,  the interference on incumbent  PUs network must be below guaranteed interference thresholds. Specifically, in this paper, we consider the problem of  the OOB emission of SU over spectrum hole affecting  the active PUs in the adjacent channels.  The resource allocation for OFDM CR is first considered in \cite{magh11}, and other heuristic and fast resource allocation methods are  proposed in\cite{magh12,magh13}. However, GFDM uses non-orthogonality of subcarriers whereas OFDM uses orthogonal ones. Thus the power allocation  problem is  completely different between GFDM and OFDM.  Thus, the   signal-to-interference-plus-noise ratio (SINR) is more complicated.  In \cite{magh14}, GFDM power allocation in underlay cognitive radio is solved via genetic algorithms. In \cite{magh15}, CR resource allocation is done by particle swarm optimization (PSO). However, although the optimization problem is not convex due to the interference on subcarriers, the dual Lagrange multiplier method is used as analytical solution in \cite{magh15}. Also, the metaheuristic approaches for non-convex optimization problems, e.g. PSO,  do not guarantee to reach to the global optimum due to the lack of any theoretical basis. To the best of our knowledge, no analytical resource allocation strategies to increase the spectral efficiency of GFDM SUs for different receiver structures have been published before.

In Fig.~\ref{system model}, we consider an SU link consisting of an SU transmitter and  SU receiver operating  over a  spectrum hole in which PUs are not present temporarily.  Also, two PUs are active in left and right channels of the spectrum hole.  Furthermore, the  interference levels from PUs to the SU link are  assumed to be negligible.  We consider GFDM system for different cases: 1) uniform and non-uniform power allocation to subcarriers, 2) different number of subsymbols and 3) two common MF and ZF  receivers. Moreover, a frequency selective slow fading channel is  considered. {We  assume that  channel state information (CSI) of all the links (e.g., SU  transmitter to receiver and SU transmitter to  PU receivers)  is available at the SU transmitter. It can estimate the CSI of SU-SU links  by using any classic channel training, estimation, and feedback mechanisms. It can estimate the CSI of SU-PU  links  by utilizing  beacon signals transmitted by PUs and by exploiting  channel reciprocity.  In this paper,  we investigate the problem of  maximizing  the SU  rate  under the constraints of maximum tolerable interference power on PU bands and maximum transmit power.  This problem is solved for the aforementioned scenarios, and GFDM is  compared with OFDM to determine the relative advantages of GFDM for CR networks.  

	\begin{figure}[]
		\centering
		\includegraphics[width=.4\textwidth]{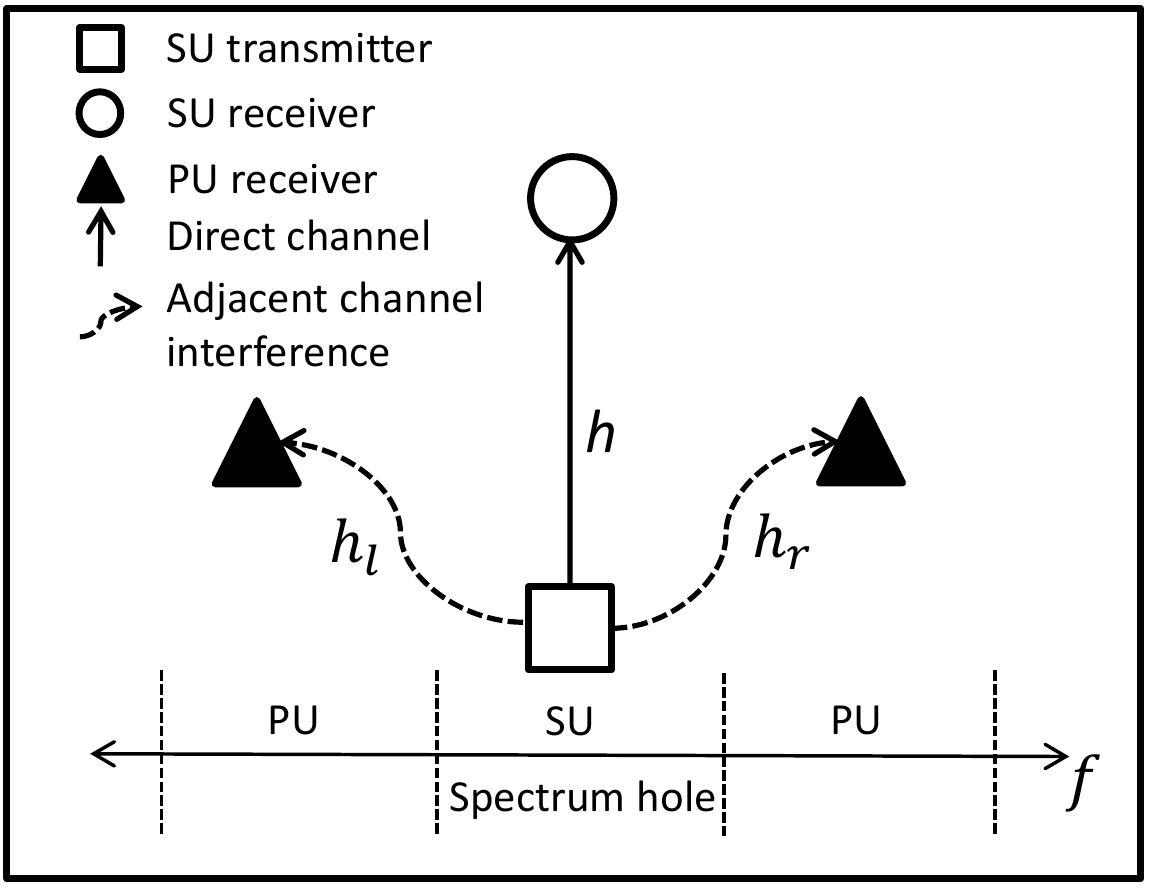}\\
		\caption {The cognitive radio model. }
		\label{system model}
	\end{figure}
    
In detail, the  contributions of this paper are follows.
\begin{itemize}
	\item For the SU link,  we  consider two different standard receiver techniques; namely  matched filter (MF) and zero forcing (ZF).  We derive their  SINR and SIR  as function of subcarrier  power allocations.  Moreover, the accuracy of the derived formulas are verified by simulations. 
	\item Adjacent channel interference (ACI) of the SU on the active users on right and left channels of the utilized spectrum hole are derived and incorporated to define the constraints of rate optimization problem. For this purpose we derive  the power spectral density (PSD) of GFDM  non-equal   subcarrier  power allocations. 
	\item The maximization problems for the total rate of SU with GFDM in a  CR network are defined.  After convexifying the  optimization problems by adding an  interference threshold, an analytical solution to the optimal subcarrier power allocations  is proposed by  utilizing the Lagrange method. 
	\item The impact of the number of subsymbols, a critical  parameter in GFDM,  on the symbol error rate (SER) performance and OOB emission of MF and ZF receivers  is investigated, and their sum rate performances are  considered.
	\item Finally,  we compare the spectral  efficiencies  of GFDM  and    OFDM. For different power allocations,  we show that GFDM achieves higher efficiency than  OFDM due to its lower OOB emission. Note that the proposed algorithm  provides high SU transmission rate  while satisfying the tolerable interference temperature constraints on PUs spectrum. These advantages are suitable  for opportunistic spectrum access in practical applications, e.g. CR TV white space (TVWS) transmission.  
	
\end{itemize}

The rest of this paper is organized as follows: the system model is presented in Section  \ref{sec:model}. The SINR and signal-to-noise ratio (SNR) of received symbols in MF and ZF receivers and their SER are derived in Section \ref{sec:bachground}. We derive the  power spectral density (PSD) of GFDM signal for non-identical  power scaling for subcarriers and obtain  the ACI on the right and left channels in Section \ref{sec:1db}. The rate optimization problems are defined and solved in Section \ref{sec:optimize}. Simulation and numerical results in Section \ref{sec:sim}  verify the accuracy of  the analytical expressions and confirm the benefits of   the  optimization results.  Finally, concluding  remarks are provided in Section \ref{sec:conclusion}.

	\section{System Model}\label{sec:model}

In our system model (Fig.~\ref{system model}), the SU uses GFDM  and the SU transmitter-receiver channel is denoted by $h$.  We assume the PUs are active in  the two adjacent channels of the spectrum hole which is used by the SU.  The gains of these two channels are denoted by ${{h}_{r}}$ and ${{h}_{l}}$. These gains will  determine the amount of ACI falls on the PUs.

In GFDM, let  $\vec{s}={{[s(0),.....,s(MK-1)]}^{T}}$ be $MK\times 1$ complex data vector with independent and identically distributed  (i.i.d) entries, which are chosen from ${{2}^{\mu }}-\text{QAM}$ complex constellation where $\mu $ is the modulation order. GFDM contains $K$ subcarriers which transmit data of $M$ time-slots. The input data vector is assigned to $M$ time-slots  according to $\vec{s}={{\left[ {{\left[ {{s}_{0}} \right]}^{T}},{{\left[ {{s}_{1}} \right]}^{T}},{{\left[ {{s}_{2}} \right]}^{T}},......,{{\left[ {{s}_{M-1}} \right]}^{T}} \right]}^{T}}$where $ \left[ {{s}_{m}} \right]={{\left[ {{s}_{m,0}},{{s}_{m,1}},{{s}_{m,2}},......,{{s}_{m,K-1}} \right]}^{T}}$.  Thus, $ {{s}_{m,k}}$ is the transmitted data symbol in   $k$-th  subcarrier of $m$-th time-slot. The GFDM signal  per frame may  be written as
	\begin{equation}\label{equation1}
    \begin{aligned}
x[n]=\sum\limits_{k=0}^{K-1}{\sum\limits_{m=0}^{M-1}{\sqrt{{{\alpha }_{k}}}{{s}_{m,k}}}{{g}_{T{{x}_{m}}}}[n]{{e}^{j2\pi \frac{(k-\frac{K-1}{2})}{K}n}}} 
\end{aligned}
	\end{equation}	
where ${{\alpha }_{k}}$ is  power allocated to  the $k$-th subcarrier,  ${{g}_{T{{x}_{m}}}}[n]={{g}_{Tx}}{{[n-mK]}_{MK}}$ is circularly shifted version of the transmitted prototype filter ${{g}_{Tx}}[n]$ and $ 0 \le n\le MK-1$. Furthermore, by representing output samples of GFDM modulator as a $MK\times 1$ vector $\vec{x}={{\left[ x[0],x[1],x[2],.....,x[MK-1] \right]}^{T}}$, one block of GFDM signal is given by $\vec{x}=\mathbf A\vec{s}$
, where $\mathbf A$ is a $MK\times MK$ modulation matrix given by ${{\left[ \mathbf A \right]}_{n,mM+k}}={{g}_{T{{x}_{m}}}}[n]{{e}^{j2\pi n\frac{(k-\frac{K-1}{2})}{K}}}$.

The signal $ x[n] $ is sent over the  wireless channel.  Given sufficiently long   CP,  perfect synchronization and knowledge of the channel impulse response, the receiver can remove  the CP. As a result, circular convolution with channel impulse response can be considered as $\vec{y}=\vec{h}\otimes\vec{x}+\vec{w}$ , where $\otimes $  denotes circular convolution and  $w$ is additive white Gaussian noise (AWGN). Then, frequency domain equalization (FDE) is used and The equalized signal is  $\vec{u}=\vec{x}+{\vec{w}_{eq}}$, where  ${\vec{w}_{eq}}=\text{IFFT}\left\{ \frac{\vec{W}}{\vec{H}} \right\}$, $\vec{W}$ and  $\vec{H}$ are the noise vector and the channel response in frequency domain, respectively. The resulted vector goes through GFDM demodulator and vector of the estimated symbols is calculated by $\vec{\widehat{s}}=\mathbf B \vec{u}$ , where $\mathbf B$ is receiver matrix. 

The receiver matrix for MF and ZF  linear GFDM receivers are equal to ${{\mathbf A}^{H}}$ and ${{\mathbf A}^{-1}}$, respectively. In the MF receiver  $ \mathbf B $  is selected to  maximizing the signal-to-noise ratio (SNR) for each symbol without considering interference. In  GFDM, however, subcarriers and subsymbols are not mutually orthogonal. Thus, the inter-subcarrier interference will limit the  performance of the MF receiver.  To eliminate it,   the linear ZF receiver can be used.  However, the ZF receiver  has the drawback of enhancing the additive noise. 

With either of these   receivers, each estimated symbol at a given subcarrier and time-slot can be written as\cite{magh7}
	 	\begin{equation}\label{equation8}
{{\widehat{s}}_{{{m}^{'}},{{k}^{'}}}}={{r}_{{{m}^{'}},{{k}^{'}}}}+{{w}_{eq,{{m}^{'}},{{k}^{'}}}}
	 	\end{equation}
where ${{w}_{eq,{{m}^{'}},{{k}^{'}}}}$ is the equivalent noise and ${{r}_{{{m}^{'}},{{k}^{'}}}}$ is equal to 

\begin{equation}\label{equation100}
\begin{aligned}
{{r}_{{{m}^{'}},{{k}^{'}}}}=\frac{1}{\sqrt{{{\alpha }_{{{k}^{'}}}}}}\sum\limits_{n=0}^{MK-1}{x[n] g_{R{{x}_{{{m}^{'}}}}}^{*}{{e}^{-j2\pi \frac{({{k}^{'}}-\frac{K-1}{2})}{K}n}}}
\end{aligned}
\end{equation} 
where ${{g}_{R{{x}}}[n]}$ is the receive filter impulse response, ${{g}_{R{{x}_{m}}}}[n]={{g}_{Rx}}{{[n-mK]}_{MK}}$  is circularly shifted version of that and $(.)^*$ denotes the conjugate operator.

%In this paper, we consider two  standard receivers -- namely ZF and MF. In both of these,  the first step is to linearly preprocess  the received signal  vector by as $ {\widehat{s}} =  \mathbf{A}u $.  In the MF receiver  $ \mathbf A $  is selected to  maximizing the signal-to-noise ratio (SNR) for each symbol without considering interference. In  GFDM, however, subcarriers and subsymbols are not mutually orthogonal. Thus, performance of the MF receiver is limited by the inter-subcarrier interference. To eliminate the inter-subcarrier interference,  the linear ZF receiver can be used with reduced complexity.

	\section{Received SINR Derivations}\label{sec:bachground}
In this section, we derived the SINR for the MF and  ZF receivers. 
	
	\subsection{ MF receiver}
	As mentioned before, the  MF receiver suffers from self-generated interference. Nevertheless, SNR per subcarrier in this receiver is maximized without considering this interference. By using  (\ref{equation1}), (\ref{equation8}), (\ref{equation100}) and substituting $m={{m}^{'}}$ and $k={{k}^{'}}$, the output of this receiver  (\ref{equation8}) can be  rewritten as			 		\begin{equation}\label{equation10}
{{\widehat{s}}_{{{m}^{'}},{{k}^{'}}}}={{s}_{{{m}^{'}},{{k}^{'}}}}+{{n}_{{{m}^{'}},{{k}^{'}}}}+{{w}_{eq,{{m}^{'}},{{k}^{'}}}^{MF}}
	 	\end{equation}
where ${{n}_{{{m}^{'}},{{k}^{'}}}}={{r}_{{{m}^{'}},{{k}^{'}}}}-{{s}_{{{m}^{'}},{{k}^{'}}}}$ is interference noise. To derive  SINR, the variance of each term is needed. First, variance of interference noise is derived as (see Appendix A).  
\begin{equation}\label{equation13}
\sigma _{{{n}_{{{m}^{'}},{{k}^{'}}}}}^{2}=\mathbb E[{{n}_{{{m}^{'}},{{k}^{'}}}}{{n}^{*}}_{{{m}^{'}},{{k}^{'}}}]=\frac{1}{{{\alpha }_{{{k}^{'}}}}}\sum\limits_{k=0}^{K-1}{{{\alpha }_{k}}}{{f}_{{{m}^{'}},{{k}^{'}}}}(k)-\overline{{{p}_{s}}} 
	 	\end{equation}      
where $\mathbb {E}[.]$ is the ensemble average operator and
\begin{equation}\label{equation101}
\begin{aligned}
{{f}_{{{m}^{'}},{{k}^{'}}}}(k)=&\sum\limits_{m=0}^{M-1}{\sum\limits_{{{n}_{1}}=0}^{MK-1}{\sum\limits_{{{n}_{2}}=0}^{MK-1}{{{\alpha }_{k}}}}\overline{{{p}_{s}}}}{{g}_{R{{x}_{m}}}}[{{n}_{1}}]
\\ &\times g_{R{{x}_{m}}}^{*}[{{n}_{2}}]g_{R{{x}_{{{m}^{'}}}}}^{*}[{{n}_{1}}]{{g}_{R{{x}_{{{m}^{'}}}}}}[{{n}_{2}}]{{e}^{j2\pi \frac{(k-{{k}^{'}})}{K}({{n}_{1}}-{{n}_{2}})}}.
\end{aligned}
\end{equation} 

 Second, the variance of equivalent noise is calculated as \cite{magh16}
 
	 	\begin{equation}\label{equation14}
\sigma _{{{w}_{eq,{{m}^{'}},{{k}^{'}}}^{MF}}}^{2}=\frac{{{N}_{0}}}{MK{{\alpha }_{{{k}^{'}}}}}{{\sum\limits_{p=0}^{MK-1}{\left| \frac{{{G}_{{{m}^{'}},{{k}^{'}}}^{MF}}[-p]}{H[p]} \right|}}^{2}}
	 	\end{equation}	 	 	
where ${{G}_{{{m}^{'}},{{k}^{'}}}^{MF}}[p]$ is frequency response of $g_{R{{x-MF}_{{{m}^{'}}}}}^{*}[n]{{e}^{-j2\pi \frac{({{k}^{'}}-\frac{K-1}{2})}{K}n}}$ , $H[p]$ is channel frequency response and ${{N}_{0}}$ is noise power density. Due to (\ref{equation13}) and (\ref{equation14}), the SINR experienced by SU receiver for MF at $k$-th subcarrier and $m$-th  time-slot after the frequency selective AWGN channel can be expressed as	
\newcommand{\snrmf}{\Gamma_{_{{{m}^{'}},{{k}^{'}}}}^{MF}}
	 	\begin{equation}\label{equation15}
        \begin{aligned}
\snrmf =&{{R}_{T}}\frac{\overline{{{p}_{s}}}}{\sigma _{{{n}_{{{m}^{'}},{{k}^{'}}}}}^{2}+\sigma _{{{w}_{eq,{{m}^{'}},{{k}^{'}}}^{MF}}}^{2}}
\\&={{R}_{T}}\frac{\overline{{{p}_{s}}}{{\alpha }_{{{k}^{'}}}}}{\sum\limits_{k=0}^{K-1}{{{\alpha }_{k}}}{{f}_{{{m}^{'}},{{k}^{'}}}}(k)-\overline{{{p}_{s}}}{{\alpha }_{{{k}^{'}}}}+{{C}_{{{m}^{'}},{{k}^{'}}}^{MF}}}
        \end{aligned}
	 	\end{equation} 	
where ${{C}_{{{m}^{'}},{{k}^{'}}}^{MF}}={{\alpha }_{{{k}^{'}}}}{\sigma _{{{w}_{eq,{{m}^{'}},{{k}^{'}}}^{MF}}}^{2}}$, ${{R}_{T}}=\frac{MK}{MK+{{N}_{CP}}}$ and  $\overline{{{p}_{s}}}=\mathbb {E}\{{{s}_{{{m}^{'}},{{k}^{'}}}}{{s}^{*}}_{{{m}^{'}},{{k}^{'}}}\}$ is average power of data symbols which for ${{2}^{\mu }}-\text{QAM}$ modulation is equal to $\overline{{{p}_{s}}}=\frac{2({{2}^{\mu }}-1)}{3}$. Furthermore, SER can be calculated for a GFDM system over frequency selective channel with MF receiver by summation of probability of symbols decoded being in error for ${{2}^{\mu }}-\text{QAM}$ as \cite{magh16}
	 		
	 	\begin{equation}\label{equation16}
	 	\begin{aligned}
{P}_{s}^{MF}=&2\left( \frac{\mu -1}{\mu MK} \right)\sum\limits_{m=0}^{M-1}{\sum\limits_{k=0}^{K-1}{\erfc\left(\sqrt{\frac{3\snrmf}{2({{2}^{\mu }}-1)}}\right)}}\\&-\frac{1}{MK}{{\left( \frac{\mu -1}{\mu } \right)}^{2}}\sum\limits_{m=0}^{M-1}{\sum\limits_{k=0}^{K-1}{\erfc^{2}\left(\sqrt{\frac{3\snrmf}{2({{2}^{\mu }}-1)}}\right)}}.
\end{aligned}
	 	\end{equation}
where $\erfc(x)$ is the complementary error function. The SER  is an important parameter of quality of service (QoS)  and (\ref{equation16}) provides the means to test  the accuracy of (\ref{equation15}).

 \subsection{ZF receiver}
	
Unlike  MF, ZF eliminates  self-generated interference, but enhances additive noise. By considering (\ref{equation8}), ZF estimated data of $m$-th time-slot in $k$-th  subcarrier can be derived as	
	 	\begin{equation}\label{equation17}
{{\widehat{s}}_{{{m}^{'}},{{k}^{'}}}}={{s}_{{{m}^{'}},{{k}^{'}}}}+{{w}_{eq,{{m}^{'}},{{k}^{'}}}^{ZF}}.
	 	\end{equation}
Due to (\ref{equation14}), the  received SNR  at $k$-th subcarrier and $m$-th time-slot given  the frequency selective and AWGN channel may  be expressed as 
\newcommand{\snrzf}{\Gamma_{_{{{m}^{'}},{{k}^{'}}}}^{ZF}}
 	 	\begin{equation}\label{equation19}
\snrzf ={{R}_{T}}\frac{\overline{{{p}_{s}}}}{\sigma _{{{w}_{eq,{{m}^{'}},{{k}^{'}}}^{ZF}}}^{2}}={{R}_{T}}\frac{\overline{{{p}_{s}}}{{\alpha }_{{{k}^{'}}}}}{C_{{{m}^{'}},{{k}^{'}}}^{ZF}}.
 	 	\end{equation}

Moreover, SER of GFDM  with ZF receiver may be given  by \cite{magh16}
	\begin{equation}\label{equation20}
	 	\begin{aligned}
{P}_{s}^{ZF}=&2\left( \frac{\mu -1}{\mu MK} \right)\sum\limits_{m=0}^{M-1}{\sum\limits_{k=0}^{K-1}{\erfc \left(\sqrt{\frac{3\snrzf} {2({{2}^{\mu }}-1)}}\right)}}\\&-\frac{1}{MK}{{\left( \frac{\mu -1}{\mu } \right)}^{2}}\sum\limits_{m=0}^{M-1}{\sum\limits_{k=0}^{K-1}{\erfc^2 \left (\sqrt{\frac{3\snrzf}  {2({{2}^{\mu }}-1)}}\right)}}.
\end{aligned}
 	 	\end{equation}	

Note  that variance of equivalent noise is derived based on receiver filter which is different for MF and ZF.

%Moreover, SER can be calculated for a GFDM system over frequency selective fading channel with ZF receiver by

 %	 	\begin{equation}\label{equation20}
%{{P}_{s}}=2\left( \frac{\mu -1}{\mu MK} \right)\sum\limits_{m=0}^{M-1}{\sum\limits_{k=0}^{K-1}{erfc(\sqrt{\frac{3SN{{R}_{m,k}}}{2({{2}^{\mu }}-1)}})}}-\frac{1}{MK}{{\left( \frac{\mu -1}{\mu } \right)}^{2}}\sum\limits_{m=0}^{M-1}{\sum\limits_{k=0}^{K-1}{erf{{c}^{2}}(\sqrt{\frac{3SN{{R}_{m,k}}}{2({{2}^{\mu }}-1)}})}}
 %	 	\end{equation}	

	\section{ Adjacent Channel Interference}\label{sec:1db}
In our model, the PUs are active in right and left adjacent channels of the spectrum hole. Thus, they will experience destructive interference because of the OOB emissions of  the SU. To study this effect, we assume frequency selective slow fading channels for PUs. However, with a sufficient cyclic prefix,  the frequency selective channel is  equivalent to  multiple flat fading channels in frequency domain. For small frequency bin, the channel frequency response is constant across each frequency bin which are denoted by ${{H}_{r}}(d)$ and ${{H}_{l}}(d)$ for right and left neighboring channel responses in each frequency bin, respectively, an example is  shown in Fig.~\ref{PSD1} for four subcarriers. The total ACI is calculated through summation of ACI on each frequency bin which is derived by multiplying the adjacent channel power (ACP) in each frequency bin by channel gains. Therefore, total ACI at right neighboring channel may be expressed as 
%Tapped delay line filter with ${{N}_{Ch}}$ independent taps is considered as a model for frequency selective channel which is written as	
 	\begin{figure}[]
		\centering
		\includegraphics[width=.40\textwidth]{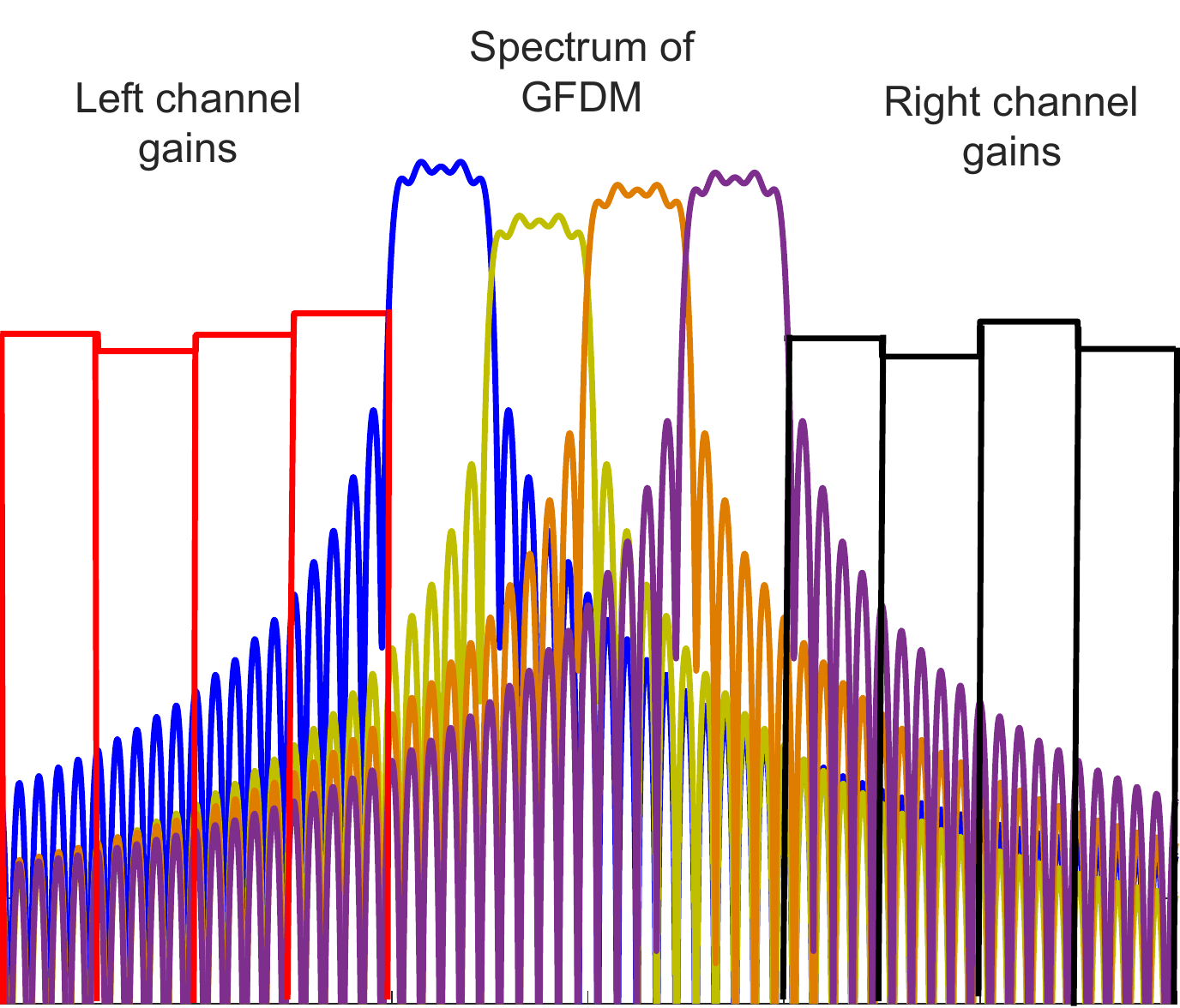}\\
		\caption {PSD of GFDM and left and right channel gains for four subcarriers. }
		\label{PSD1}
	\end{figure}

 	 	\begin{equation}\label{equation23}
{{P}_{ACI}}=\sum\limits_{d=K+1}^{2K}{{{P}_{AC}}({{f}_{d}})}{{H}_{r}}(d-K)
 	 	\end{equation}	 
where ${{P}_{AC}}({{f}_{d}})$ is ACP in frequency interval $[{{f}_{d}}-1/(2{{T}_{s}}),{{f}_{d}}+1/(2{{T}_{s}})]$ where ${{f}_{d}}=\frac{K+2d+1}{2{{T}_{s}}}$  is center of each frequency interval ,  ${{T}_{s}}$  is one time-slot duration and $d$ is the
index of frequency bin. To find ACP in each frequency bin, the PSD of signal is derived as (see Appendix B)

 	 	\begin{equation}\label{equation26}
{{S}_{xx}}(f)=\frac{\overline{{{p}_{s}}}}{M{{T}_{s}}}\sum\limits_{k=0}^{K-1}{{{\alpha }_{k}}{{S}_{GG}}(f-}\frac{(k-\frac{K-1}{2})}{{{T}_{s}}}) 
 	 	\end{equation} 	 	
where ${{S}_{GG}}(f)=\sum\limits_{m=0}^{M-1}{{{\left| {{G}_{T{{x}_{m}}}}(f) \right|}^{2}}}$ and ${{G}_{T{{x}_{m}}}}(f)$ is frequency response of each filter. According to (\ref{equation23}) and (\ref{equation26}), ACI in right adjacent channel can be derived as 

 	 	\begin{equation}\label{equation27}
{{P}_{r}}=\sum\limits_{k=0}^{K-1}{{{\alpha }_{k}}{{T}_{r}}(k)} 
 	 	\end{equation}	
 where 
 \begin{equation}\label{equation260}
 {{T}_{r}}(k)=\frac{\overline{{{p}_{s}}}}{MT_{s}}\sum\limits_{d=K+1}^{2K}{H_{r}(d-K)}\int\limits_{f_{d}-1/(2T_{s})}^{f_{d}+1/(2T_{s})} {{S}_{GG}}(f-\frac{(k-\frac{K-1}{2})}{{{T}_{s}}}df.
 \end{equation}
 
 Similarly,  ACI in left adjacent channel is calculated by 	 	 	
\begin{equation}\label{equation28}
{{P}_{l}}=\sum\limits_{k=0}^{K-1}{{{\alpha }_{k}}{{T}_{l}}(k)} 
 	 	\end{equation} 	
        
where

 \begin{equation}\label{equation260}
{{T}_{l}}(k)=\frac{\overline{{{p}_{s}}}}{MT_{s}}\sum\limits_{d=K+1}^{2K}{{{H}_{l}}(d-K)}\int\limits_{-{{f}_{d}}-1/(2{{T}_{s}})}^{-{{f}_{d}}+1/(2{{T}_{s}})} {{S}_{GG}}(f-\frac{(k-\frac{K-1}{2})}{{{T}_{s}}}df.
 \end{equation}
These two derived powers must be below the acceptable interference thresholds of the left and right PU channels.  This constraint will be incorporated into the  rate optimization problem subsequently.

 	\section{Problem Formulation}\label{sec:optimize}	 	
We next formulate and solve  the maximization of the total transmission rate of the SU under interference constraint for both MF and ZF receivers. 	Technically, we aim to find the optimal set of power allocations $(\alpha_0, \alpha_1,\ldots, \alpha_{K-1} )$. 

\subsection{MF receiver}
With the MF receiver, according to (\ref{equation15}),(\ref{equation27}) and (\ref{equation28}), the rate optimization problem can be formulated as

 	 	\begin{equation}\label{equation29}
 \underset{\begin{smallmatrix} 
	\alpha_{k} \\ 
	0\le k\le K-1 
	\end{smallmatrix}}{\mathop{\max }}\,\sum\limits_{{{k}^{'}}=0}^{K-1}{\sum\limits_{{{m}^{'}}=0}^{M-1}{\log (1+\snrmf)}}  
 	 	\end{equation} 	
        
 	 	\begin{equation}\label{equation103}
 s.t.\quad\sum\limits_{k=0}^{K-1}{{{\alpha }_{k}}<{{\alpha }_{\max }}} 
 	 	\end{equation} 	
         	 	\begin{equation}\label{equation104}
\qquad \sum\limits_{k=0}^{K-1}{{{\alpha }_{k}}{{T}_{r}}(k)}<{{Q}_{r}} 
 	 	\end{equation} 
        
         	 	\begin{equation}\label{equation105}
\qquad \sum\limits_{k=0}^{K-1}{{{\alpha }_{k}}{{T}_{l}}(k)}<{{Q}_{l}} 
 	 	\end{equation} 
where ${{Q}_{r}}$ and ${{Q}_{l}}$ are the interference temperature limits for the right and left adjacent channels, respectively. Furthermore, ${{\alpha }_{\max }}$ is the maximum total power that is divided between $K$ subcarriers.  However, because  $\Gamma_{_{{{m}^{'}},{{k}^{'}}}}^{\text{MF}}$ is a rational expression of $\alpha_k$'s, the  objective function  (\ref{equation29}) is not convex. To overcome the non-convexity of objective function, we introduce an additional constraint to original problem \cite{magh19} which is given by 	 	 	

 	 	\begin{equation}\label{equation30}
\sum\limits_{k=0}^{K-1}{{{\alpha }_{k}}}{{f}_{{{m}^{'}},{{k}^{'}}}}(k)-\overline{{{p}_{s}}}{{\alpha }_{{{k}^{'}}}}<{{Q}_{n}}\quad\forall {{m}^{'}},{{k}^{'}}.
 	 	\end{equation}

As shown in (\ref{equation10}) , due to self-generated interference in MF receiver, extracted symbols contain interference noise with which the total power of these terms have been calculated in (\ref{equation13}). Thus, inequality constraint (\ref{equation30}) can be interpreted as the sum of  self-generated interference in each time-slot of each subcarrier  being less than the self-interference noise threshold  ${{Q}_{n}}$.  By setting this threshold  ${{Q}_{n}}$ appropriately,  we can improve the system performance. The resulting  convex optimization problem is solved by utilizing the method of Lagrange multipliers to encapsulate all  the constraints: 
\begin{equation}\label{equation33}
\begin{aligned}
& L(\alpha ,\gamma )=\sum\limits_{{{k}^{'}}=0}^{K-1}{\sum\limits_{{{m}^{'}}=0}^{M-1}{\log (1+\frac{\overline{{{p}_{s}}}{{\alpha }_{{{k}^{'}}}}}{{{Q}_{n}}+{{C}_{{{m}^{'}},{{k}^{'}}}^{MF}}})}}\\&+\sum\limits_{{{k}^{'}}=0}^{K-1}{\sum\limits_{{{m}^{'}}=0}^{M-1}{{{\gamma }_{{{m}^{'}}+M{{k}^{'}}}}}}({{Q}_{n}}-\sum\limits_{k=0}^{K-1}{{{\alpha }_{k}}}{{f}_{{{m}^{'}},{{k}^{'}}}}(k)+\overline{{{p}_{s}}}{{\alpha }_{{{k}^{'}}}}) \\ 
& +{{\gamma }_{MK}}({{\alpha }_{\max }}-\sum\limits_{k=0}^{K-1}{{{\alpha }_{k}}})+{{\gamma }_{MK+1}}({{Q}_{r}}-\sum\limits_{k=0}^{K-1}{{{\alpha }_{k}}{{T}_{r}}(k)})\\&+{{\gamma }_{MK+2}}({{Q}_{l}}-\sum\limits_{k=0}^{K-1}{{{\alpha }_{k}}{{T}_{l}}(k)}) \\ 
\end{aligned}
 	 	\end{equation}
        
                \newcounter{storeeqcounter}
\newcounter{tempeqcounter}
\begin{figure*}[!t]
\normalsize
\setcounter{tempeqcounter}{\value{equation}} % temp store of current value
\begin{equation}{}
\setcounter{storeeqcounter}{\value{equation}} % number of this equation
{{\alpha }_{k}}={{\left[ \frac{M}{\left[ \sum\limits_{{{k}^{'}}=0}^{K-1}{\sum\limits_{{{m}^{'}}=0}^{M-1}{{{\gamma }_{{{m}^{'}}+M{{k}^{'}}}}}}({{f}_{{{m}^{'}},{{k}^{'}}}}(k)-\overline{{{p}_{s}}}) \right]+{{\gamma }_{MK}}+{{\gamma }_{MK+1}}{{T}_{r}}(k)+{{\gamma }_{MK+2}}{{T}_{l}}(k)}-\frac{{{Q}_{n}}+{{C}_{0,k}^{MF}}}{\overline{{{p}_{s}}}} \right]}^{+}}
\label{eq:floatingeq}
\end{equation}
\setcounter{equation}{\value{storeeqcounter}} % restore correct value
\hrulefill
\vspace*{4pt}
\end{figure*}
where ${\gamma}_{j}$ , $j=0,...,MK+2 $, are Lagrange  multipliers. Due to the  standard convex form of this problem,  the Karush-Kuhn-Tucker (KKT) conditions,  which are the first order necessary and sufficient conditions for optimality,  yield the optimal solution. Thus, the optimal power allocated to each subcarrier as a function of the Lagrange multipliers  is obtained as \eqref{eq:floatingeq}, where ${[x]^{+}}=\max (0,x)$ and $k$ is the subcarrier index. From the  Lagrangian duality, (\ref{equation33}) should be minimized on Lagrangian multipliers.  Thus, they can updated by using  the sub-gradient method. The subgradient update equations are given by

 	 	\begin{equation}\label{equation35}
\begin{aligned}
& {{\gamma }_{{{m}^{'}}+M{{k}^{'}}}}(t+1)={{\left[ {{\gamma }_{{{m}^{'}}+M{{k}^{'}}}}(t)+\zeta (\sum\limits_{k=0}^{K-1}{{{\alpha }_{k}}}{{f}_{{{m}^{'}},{{k}^{'}}}}(k)-\overline{{{p}_{s}}}-{{Q}_{n}}) \right]}^{+}}\\ 
& {{\gamma }_{MK}}(t+1)={{\left[ {{\gamma }_{MK}}(t)+\zeta (\sum\limits_{k=0}^{K-1}{{{\alpha }_{k}}-{{\alpha }_{\max }}}) \right]}^{+}} \\ 
& {{\gamma }_{MK+1}}(t+1)={{\left[ {{\gamma }_{MK+1}}(t)+\zeta (\sum\limits_{k=0}^{K-1}{{{\alpha }_{k}}{{T}_{r}}(k)}-{{Q}_{r}}) \right]}^{+}} \\ 
& {{\gamma }_{MK+2}}(t+1)={{\left[ {{\gamma }_{MK+2}}(t)+\zeta (\sum\limits_{k=0}^{K-1}{{{\alpha }_{k}}{{T}_{l}}(k)}-{{Q}_{l}}) \right]}^{+}} \\ 
\end{aligned}
 	 	\end{equation}
where $\zeta$ denotes the positive step size. Thus, the entire iterative process for solving this  rate optimization problem is given in Algorithm. 1. Due to the convexity of our problem, the duality gap is zero and the  proposed algorithm is optimal. We should note that if equal power for subcarriers are considered,  we can set  $\alpha_{k}=\alpha$. Thus, the optimum amount of the power which satisfies all constraints could be derived as
\begin{equation}\label{equation32}
{{\alpha }_{opt}}=\min \left\{ \frac{{{Q}_{n}}}{\sum\limits_{k=0}^{K-1}{{{f}_{{{m}^{'}},{{k}^{'}}}}(k)}-\overline{{{p}_{s}}}},\frac{{{\alpha }_{\max }}}{K},\frac{{{Q}_{r}}}{\sum\limits_{k=0}^{K-1}{{{T}_{r}}(k)}},\frac{{{Q}_{l}}}{\sum\limits_{k=0}^{K-1}{{{T}_{l}}(k)}} \right\}.
 	 	\end{equation}
This solution is the just the power level that will not violate all the constraints. 

	\begin{algorithm}[t]{}
		\caption{Rate optimization}\label{alg:dinkel}
		1: Initialize the maximum number of iteration ${{I}_{\max }}$ and convergence condition ${{\varepsilon }_{\gamma }}$ \\
		2: set $t\leftarrow 1$ \\
		3: \textbf {do while} $\sum\limits_{i=1}^{K-1}{\left| {{\alpha }_{k}}(t)-{{\alpha }_{k}}(t-1) \right|>{{\varepsilon }_{\gamma }}}$  and $t<{{I}_{\max }}$ \\
		4: $t\leftarrow t+1$ \\
		5: Obtain ${\alpha}_{k}$ by using derived formula   \\
		6: Update the Lagrange multipliers  \\
		7: \textbf {end do}\\
		8: \textbf{return}
	\end{algorithm}

\subsection{ZF receiver}
	
Like the MF receiver, the ZF  rate  maximization problem  is formulated as
 	 	\begin{equation}\label{equation36}
\underset{\begin{smallmatrix} 
	\alpha_{k} \\ 
	0\le k\le K-1 
	\end{smallmatrix}}{\mathop{\max }}\,\sum\limits_{{{k}^{'}}=0}^{K-1}{\sum\limits_{{{m}^{'}}=0}^{M-1}{\log (1+\snrzf)}} \\ 
 	 	\end{equation}	
        
   	 	\begin{equation}\label{equation39}\nonumber
s.t.~~~~~~~ (20), (21), (22).
  	 	\end{equation}       
Since this problem is convex, the  Lagrangian  for this optimization problem is obtained as
\begin{equation}\label{equation38}
\begin{aligned}
& L(\alpha ,\gamma )=\sum\limits_{{{k}^{'}}=0}^{K-1}{\sum\limits_{{{m}^{'}}=0}^{M-1}{\log (1+\frac{\overline{{{p}_{s}}}{{\alpha }_{{{k}^{'}}}}}{{C}_{{{m}^{'}},{{k}^{'}}}^{ZF}})}}\\&+{{\gamma }_{0}^{'}}({{\alpha }_{\max }}-\sum\limits_{k=0}^{K-1}{{{\alpha }_{k}}})+{{\gamma }_{1}^{'}}({{Q}_{r}}-\sum\limits_{k=0}^{K-1}{{{\alpha }_{k}}{{T}_{r}}(k)}) \\ 
& +{{\gamma }_{2}^{'}}({{Q}_{l}}-\sum\limits_{k=0}^{K-1}{{{\alpha }_{k}}{{T}_{l}}(k)}) \\ 
\end{aligned}
 	 	\end{equation}	
where ${\gamma}_{i}^{'}$ , $i=0,1,2 $, are Lagrangian multipliers.  Based on KKT conditions, the optimal power allocated to each subcarrier is calculated as
  	 	\begin{equation}\label{equation39}
{{\alpha }_{k}}={{\left[ \frac{M}{{{\gamma }_{0}^{'}}+{{\gamma }_{1}^{'}}{{T}_{r}}(k)+{{\gamma }_{2}^{'}}{{T}_{l}}(k)}-\frac{{{C}_{0,k}^{ZF}}}{\overline{{{p}_{s}}}} \right]}^{+}}.
  	 	\end{equation}	
  	 	
The subgradient equations for updating the lagrangian coefficients in each iteration can be derived as

  	 	\begin{equation}\label{equation40}
\begin{aligned}
& {{\gamma }_{0}^{'}}(t+1)={{\left[ {{\gamma }_{0}^{'}}(t)+\zeta (\sum\limits_{k=0}^{K-1}{{{\alpha }_{k}}-{{\alpha }_{\max }}}) \right]}^{+}} \\ 
& {{\gamma }_{1}^{'}}(t+1)={{\left[ {{\gamma }_{1}^{'}}(t)+\zeta (\sum\limits_{k=0}^{K-1}{{{\alpha }_{k}}{{T}_{r}}(k)}-{{Q}_{r}}) \right]}^{+}} \\ 
& {{\gamma }_{2}^{'}}(t+1)={{\left[ {{\gamma }_{2}^{'}}(t)+\zeta (\sum\limits_{k=0}^{K-1}{{{\alpha }_{k}}{{T}_{l}}(k)}-{{Q}_{l}}) \right]}^{+}}. \\ 
\end{aligned}
  	 	\end{equation}
  As before,  the optimization problem  (\ref{equation36}) is solved with  Algorithm 1.
 Same as previous section, the optimum problem for uniform power allocation is derived as

 	 	\begin{equation}\label{equation37}
{{\alpha }_{opt}}=\min \left\{ \frac{{{\alpha }_{\max }}}{K},\frac{{{Q}_{r}}}{\sum\limits_{k=0}^{K-1}{{{T}_{r}}(k)}},\frac{{{Q}_{l}}}{\sum\limits_{k=0}^{K-1}{{{T}_{l}}(k)}} \right\}.
 	 	\end{equation}  
        
Note  that all the derived  analytical formulas are  general and hold  for arbitrary   signaling alphabet, numbers of  subcarriers and subsymbols and any  type of prototype filter. Indeed, in all cases, our proposed algorithms for MF and ZF can provide optimal  power allocation to maximize the rate of SU.

	\section{simulation and Numerical results}\label{sec:sim}
	
In this section, the derived SERs and PSD  are validated with simulations and optimization results are presented. The parameters of GFDM  and channel are shown in Table \ref{table}. The Averaged periodogram algorithm with 50$\%$ overlap and Hanning window is used to estimate the PSD and length of FFT (fast Fourier transform)  is set to 65536.  Monte carlo simulation  uses  1000 GFDM symbols per run. Moreover, the value of threshold is equal to 0.0001 and $\alpha_{max}=55\;$dBm.  In Section \ref{sec1}, the  SER  of (\ref{equation16}) and (\ref{equation20}) are verified by simulations.  The accuracy of derived PSD for GFDM modulated signal (\ref{equation26}) is validated by simulation.  In Section \ref{sec2},  the performance of proposed Algorithm 1 for both  GFDM and OFDM with uniform and non-uniform power allocation to subcarriers is  evaluated. 

	\begin{table}
				\caption{GFDM, channel and system parameters}				
				\begin{tabular}{l c}           
					%\begin{tabular}{ |p{3.5cm}||p{2cm}|p{2cm}|p{2.5cm}|  }
					\toprule                    
                    \cmidrule(r){1-2}
					parameter & Value \\
					\midrule
					Mapping& $16-QAM$ \\
					
					Filter type& $Raised-cosine$ \\
					
					Roll-off factor& $0.15$ \\
					
					Symbol duration ($T_{s}$)& 33.3 $\mu s$ \\
					
					Number of subcarriers ($K$)& 64 \\
					
					Number of sub symbols ($M$)& 5,15 \\
					
					Subcarrier spacing ($\frac{1}{T_{s}}$)&30 $KHz$\\
					
					Signal Bandwidth ($\frac{K}{T_{s}}$) &1.92 $MHz$\\
					
					CP length ($N_{cp} $) & 10\\
					
					Channel length ($N_{ch} $) & 10 \\
					
					Variance of each tap & ${{({{10}^{\frac{-i}{{{N}_{ch}}-1}}})}_{i=0,...,{{N}_{ch}}-1}}$  \\
					
					Number of OFDM subcarriers  & 64 \\
				    \bottomrule
				\end{tabular}
				\label{table}
			\end{table}		
				
	\subsection{Verification of derived results}\label{sec1}
 
In Fig.~\ref{SER}, the SER  of GFDM with MF or  ZF receivers are compared with that of  OFDM.  $E_{s}$ is average transmit power which for GFDM is equal to $E_{s}=\frac{\overline{{{p}_{s}}}}{K}\sum\limits_{k=0}^{K-1}{{{\alpha }_{k}}}$ and $ N_{0}$ is normalized to one. On the one hand, as can be seen, the simulation results verify the analytical derivation of SER for MF and ZF receivers in (\ref{equation16}) and (\ref{equation20}), respectively. Therefore, we can conclude that the derived formulas for SINR of MF (\ref{equation15}) and SNR of ZF (\ref{equation19}) are  accurate. Also, with GFDM,  ZF outperform   MF,  which has performance  approximately near that of  OFDM. On the other hand, more  subsymbols degrade  the SER performance of both MF and ZF receivers. This degradation  is due to noise enhancement in both receivers when the number of subsymbols is increased. Note that $\alpha_{max}$ has been chosen according to Fig.~\ref{SER}.

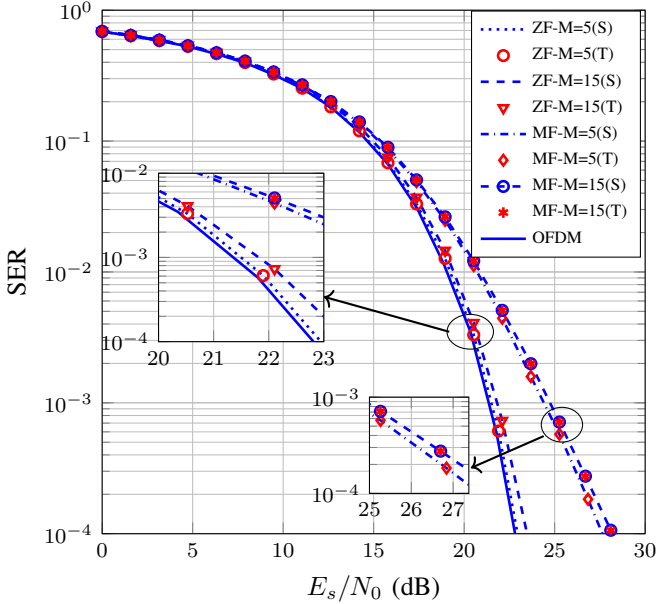
\begin{figure}
\centering
\input{SER.tex}
		\caption {SER of GFDM (with MF and ZF receivers) and of  OFDM. Legend: S=simulation and   T=theory.}
		\label{SER}
\end{figure}

	%\begin{figure}[]
	%	\centering
	%	\includegraphics[width=.5\textwidth]{SER.pdf}\\
	%	\caption {SER performance of GFDM (with MF and ZF receivers) and OFDM modulation by considering different number of subsymbols. S and T indicate the simulation and theory results, respectively. }
	%	\label{SER}
	%\end{figure}

\begin{figure}
\centering
\input{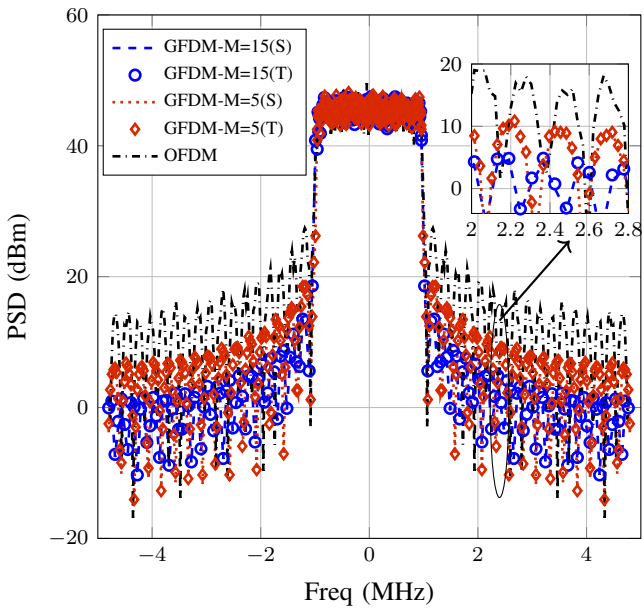}
		\caption {PSDs of GFDM and  OFDM signals.}
		\label{PSD}
\end{figure}

%	\begin{figure}[]
%		\centering
%		\includegraphics[width=.5\textwidth]{PSD1.pdf}\\
%		\caption {PSD of GFDM modulated signal with two different number of subsymbols in compare with OFDM modulated signal}
%		\label{PSD}
%	\end{figure}

Fig.~\ref{PSD} shows the PSD of GFDM with different numbers of subsymbols is compared with that of OFDM. A set of random power allocations for subcarriers is generated and has been utilized for all three scenarios.  As can be seen, the simulation results verify the derived PSD  (\ref{equation26}). Moreover,  the increased  number of subsymbols decreases the PSD rapidly on  adjacent channels. Indeed, OOB emission  decreases by increasing the number of GFDM subsymbols. This result is in contrast with effect of increasing number of subsymbols on SER performance. Thus, although more  subsymbols decreases the OOB emission,  the SER also degrades.  The SER is affected due to  in-band noise which influences the transmission rate while OOB emission  indicates the out-of-band noise which plays a main role in the CR constraints. In the next section, we investigate the trade-off between these two  effects of increasing the number of subsymbols. 
\begin{figure}
\centering
\input{Qn.tex}
		\caption {Capacity of the GFDM with MF receiver versus $Q_{n}$ for three constant interference power constraints.  }
		\label{Qn}
\end{figure}
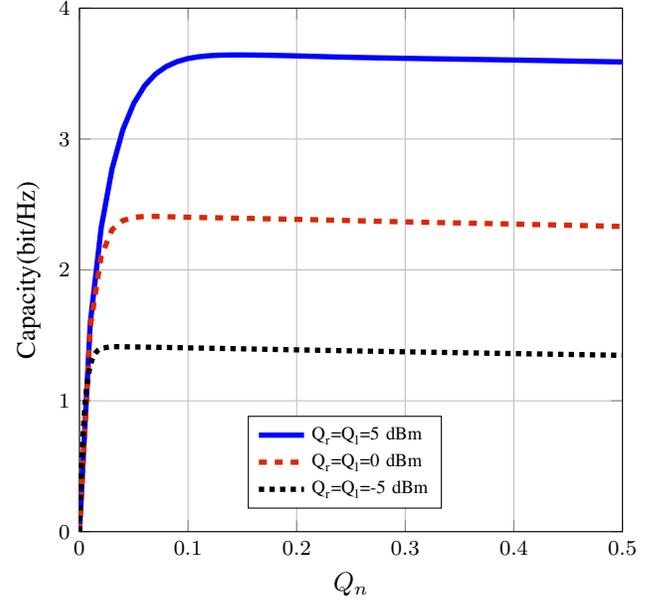

%	\begin{figure}[]
%		\centering
%		\includegraphics[width=.5\textwidth]{Qn.pdf}\\
%		\caption {Capacity of the GFDM with MF receiver versus different values of $Q_{n}$ for three constant interference power constraints.  }
%		\label{Qn}
%	\end{figure}

	\begin{figure*}[t]
		\begin{subfigure}[]{.5\textwidth}
        \input{MF.tex}
			%\includegraphics[width=1\textwidth]{ZF.pdf}\\[-10pt]
			%\textwidth
			\centering
			\caption{\fontsize{8}{8}\selectfont $MF-GFDM$}
			%\caption{\small AM/AM \& AM/PM}
			%\captionsetup[subfigure]{justification=centering}%\vspace{-0.5cm}
			\label{MF}
		\end{subfigure} %\hspace{5cm}%
		\begin{subfigure}[]{.5\textwidth}
			%\includegraphics[width=1\textwidth]{MF.pdf}\\[-10pt]
			%\textwidth
            \input{ZF.tex}
			\centering
			\caption{\fontsize{8}{8}\selectfont$ZF-GFDM$}
			%\captionsetup[subfigure]{justification=centering}%\vspace{-0.5cm}
			\label{ZF}
		\end{subfigure}
		\caption {Total sum rate of the SU of GFDM with MF and ZF receivers in compare with OFDM.}
		\label{capacityMFZF}
	\end{figure*}
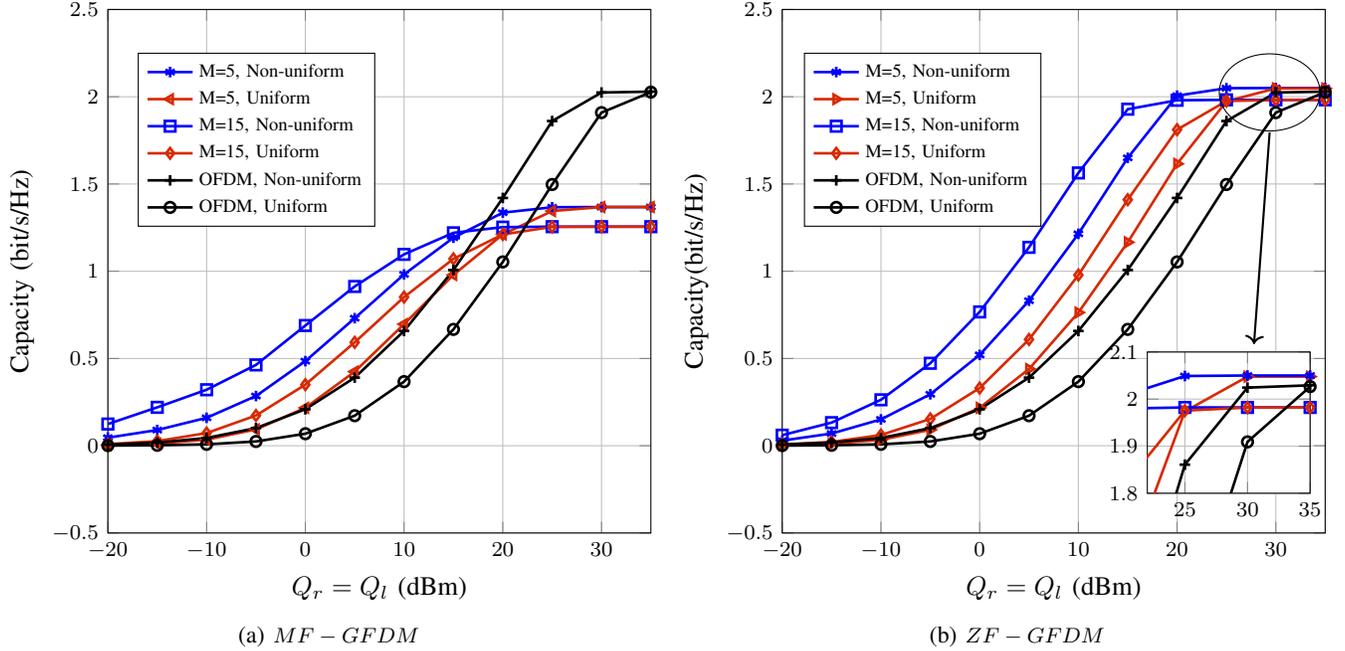

\begin{center}
\begin{table*}{}
	\caption{ Optimum values of $Q_{n}$ extracted for different interference power constraints by considering two different number of subsymbols .}
	\centering
	\begin{tabular}{ |c|c|c|c|c|c|c|c|c|c|c|  }
		%\begin{tabular}{ |p{3.5cm}||p{2cm}|p{2cm}|p{2.5cm}|  }
		\hline
		Subsymbols & -20\;dBm & -15 \;dBm & -10 \;dBm & -5 \;dBm & 0 \;dBm & 5 \;dBm & 10 \;dBm & 15 \;dBm & 20 \;dBm & 25 \;dBm  \\
		\hline
		$M=5$ & 0.0022 & 0.0044 & 0.0155 & 0.03 & 0.07 & 0.15 & 0.34 & 0.7 & 1.435 & 1.435    \\
		\hline
		$M=15$ & 0.0133 & 0.0233 & 0.04 & 0.091 & 0.1933 & 0.47 & 1.02 & 2.11 & 2.11 & 2.11  \\
		\hline
	\end{tabular}
	\label{table2}
\end{table*}{}		
\end{center}

	\subsection{Optimization Results}\label{sec2}
	
In this part, we evaluate the performance of GFDM and OFDM  systems subject to the CR constraints.  As the first step, we investigate the effect of the self-interference threshold  $Q_{n}$ on the transmission rate to find the best value of $Q_{n}$. Since this value helps us to convert the optimization problem into convex form, determining the right value  so important. After that,  the rate optimization problem is solved by using the Lagrangian method. Finally, the simulation results are presented and the performance of the proposed algorithm is compared between GFDM system with ZF and MF receivers and OFDM system and the influence of number of subsymbols is considered as well.

As  mentioned, the amount of self-interference, considered as an extra constraint and transformed the problem into convex optimization problem, has an impact on the transmission rate of  the SU. To evaluate this effect and find the optimal value, we solve the optimization problem for  the fixed value of interference power constraint. We  sweep  $Q_{n}$ over a range  to find the optimum one which maximizes the total transmission rate. Fig.~\ref{Qn} shows the transmission rate versus value of $Q_{n}$ for three amounts $Q_{r}$ and $Q_{l}$ (5\;dBm, 0\;dBm and -5\;dBm) where the number of subcarriers and subsymbols are set to 64 and 5, respectively. As expected, this power allocation problem has an optimum value. By utilizing the same procedure, for two number of subsymbols ($M=5$ and $M=15$) with different values of interference power constraint, the optimization problem is solved and the optimum values of $Q_{n}$ are given in Table\ref{table2}. In the following, these fixed value of $Q_{n}$ is chosen for the rate optimization problem.

Fig.~\ref{capacityMFZF} represents the results of the rate optimization problems for GFDM and OFDM systems. The achievable transmission rate for GFDM system with MF and ZF receivers are shown in Fig.~\ref{MF} and Fig.~\ref{ZF}, respectively, in which the results contain uniform and non-uniform power allocations to subcarriers. As expected, in both cases of MF and ZF, non-uniform power allocation has better performance in compare with allocating equal power to all subcarriers. On the one hand, in Fig.~\ref{MF}, when the interference power constraint is not dominant and the OOB emission does not cause any problem in CR e.g.$Q_{r}=Q_{l}=30$\;dBm, due to the non orthogonality of GFDM and self-interference, OFDM system should achieves higher rate. Similarly, although, the SER performance of ZF receiver is lower than OFDM, the transmission rate of that is higher than OFDM due to the inserting one CP to each GFDM frame including symbol of M time-slots. Now when the amount of interference power constraints decrease and become dominant, where is sufficient for the CR system, the GFDM system in both case of MF and ZF receivers achieves higher rates which is caused by lower OOB radiation of GFDM system in compare with OFDM system. On the other hand, we investigate the impact of number of subsymbols for GFDM system with both MF and ZF receivers. we can conclude that in low amount of interference power constraint, GFDM system with both type of receivers transmits higher rate in case of \(M=15 \) than \( M=5\) due to the lower OOB emission and is declared in Fig.~\ref{PSD}. But, since the SER performance of GFDM system is decreased by increasing the number subsymbols, when the interference power constraint is not dominant, the achievable rate in case of \(M=5\)  is higher than \(M=15.\) Consequently, in the CR system, the GFDM system with MF and ZF receivers transmits higher rate in compare with OFDM system in both case of uniform and non-uniform power allocation to subcarriers. Also, increasing the number of subsymbols can help us to achieve higher rate.

To compare the performance of ZF and MF receivers, the total transmitted rates of GFDM system for \(M=5\) subsymbols are illustrated in Fig.~\ref{joft}. When the interference power constraint is not dominant, because the  SER performance  ZF receiver is better than that of  MF receiver, the former  achieves higher transmission rate. But, when the interference power constraint decreases, the rate of MF receiver exceeds that of the ZF receiver. This phenomena  is caused by the  decreasing  amount of interference appearing in  SINR  (\ref{equation15}) when the amount of powers allocated to subcarriers decrease. Since  MF receiver maximizes the SNR without considering the interference, when the interference can be neglected, the MF receiver has the best performance. Consequently, when the interference power constraint is non-dominant, the system with ZF receiver achieves a  higher rate. But, by decreasing the interference power constraint and eliminating the interference, the system with MF receiver achieves  higher transmission rate than the one with ZF receiver.

	\begin{figure}[]
		\centering
        \input{Joft.tex}
		\caption {Total sum rate of GFDM with ZF and MF receivers for $M=5$. }
		\label{joft}
	\end{figure}
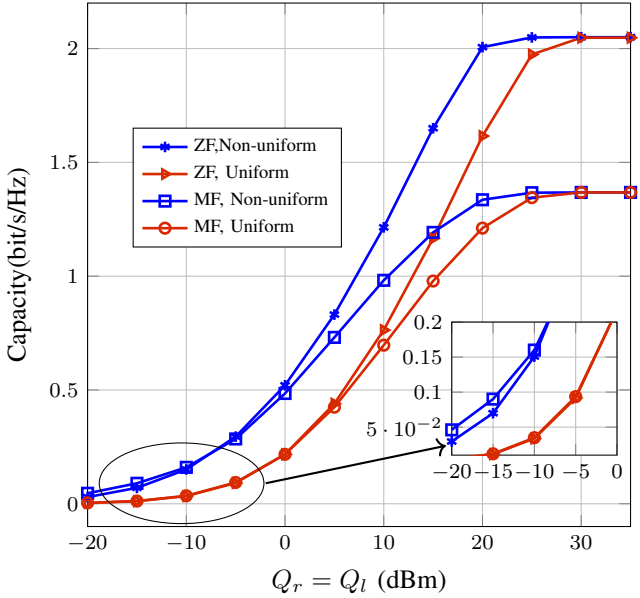

	\begin{figure}[]
		\centering
        \input{power.tex}
		\caption {Total transmitted power versus different power interference constraints.}
		\label{Power}
	\end{figure}
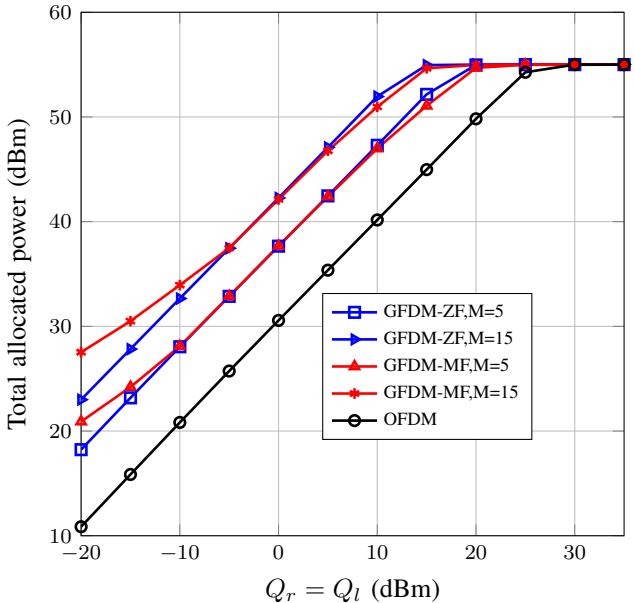

Fig.~\ref{Power} shows the total optimized transmit power versus the interference power constraint. GFDM with ZF and MF  receivers achieves higher transmit power than that  with OFDM. The reason is that the interference power constraint becomes dominant in GFDM  later than in OFDM,  which is due to lower OOB emission of GFDM system. On the other hand, more  subsymbols leads to decrease OOB emission. Thus, in case of \(M=15,\) higher power is transmitted in compare with \(M=5.\) All of these results confirm the previous results in which utilizing the GFDM system with higher number of subsymbols in lower interference power constraint can satisfy the CR system demand. Moreover, when the interference power constraint decreases,  higher transmit power is achieved by the system with the  MF receiver in comparison with ZF receiver.  This observation also agrees with  Fig.~\ref{joft}.  

	\begin{figure}[]
		\centering
        \input{tad.tex}
		\caption {Interference power versus different values of $Q_{r}$.}
		\label{tadakhol}
	\end{figure}
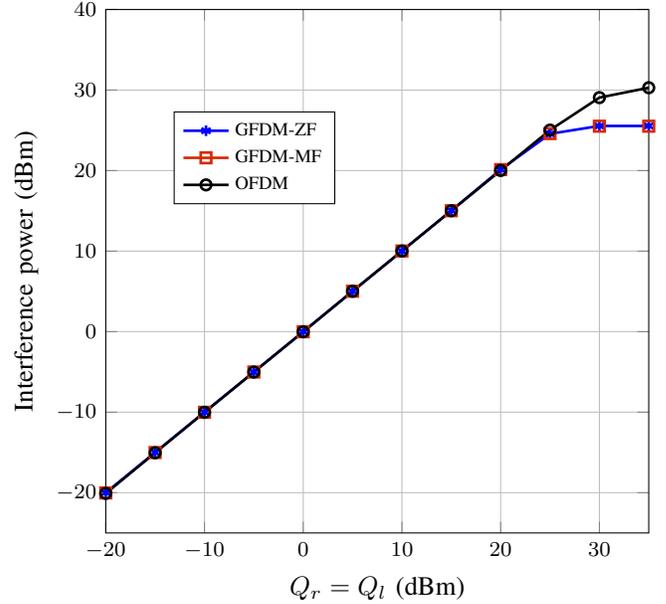

Fig.~\ref{tadakhol} represents the average interference power based on power allocation results versus interference power which contains the GFDM and OFDM systems. Both MF and ZF receivers are considered for GFDM system. As expected, the average interference power is approaching to specific value of interference power constraint. This figure shows that the optimization  realizes the interference level and not exceed the allowed value which is determined for interference power constraint. 
%Thus, the convergence of the iterative algorithm is confirmed.	  

\section{Conclusion}\label{sec:conclusion}
	
This paper  evaluated the performance of a GFDM based CR network where   unlicensed secondary  users  access the primary spectrum.  Specifically, we considered a secondary user transmitting  on a spectrum hole whose left and right adjacent bands are occupied by  primary users.  To  constrain  the ACI on the left and right channels, we derived the PSD of GFDM signal with the non-equal  power subcarriers. 
To determine  the sum rate of the SU, we derived  the SINR and SNR for  MF and ZF receivers, respectively. The SER and GSD expressions were validated over by  simulations over frequency selective fading and AWGN channels.  The  simulation results also showed that  the SER performance  degrades  and OOB leakage decreases when the number of subsymbols increases. 

We also  optimized the SU rate subject the ACI constraints  on left and right channels and maximum total power. In MF case, by adding the new constraint on limitation of self-generated interference the problem was converted to convex form. Also, the maximum self-generated interference limit of this constraint which maximizes the sum rate of the SU was extracted by simulation. The resuting  rate optimization problems  were solved by using the Lagrangian method. We compared  the total transmitted rate of the SU, utilizing GFDM with MF and ZF receivers, with OFDM for both uniform and non-uniform power allocations to subcarriers.

The simulations show that the total rate of SU for our proposed power allocation algorithm is significantly higher than uniform power allocation in all cases and that GFDM achieves higher data throughput than  OFDM in a  CR network where the ACI constraints are dominant. In this case, MF receiver can achieve more total rate compared with ZF receiver. But  when the ACI constraints are not dominant, the ZF receives achieves better  SER and data rates in comparison  with MF. In this case,  OFDM performs roughly equal to  ZF and better than MF. 
 %On the other hand, due to decreased OOB emission by increasing the number of subsymbols in GFDM, higher total rate could be transmitted by boosting number of subsymbols, which is shown by simulation results. Furthermore, the amount of total allocated power confirms our aforementioned results.
 Consequently,  GFDM  based  CR nodes  with MF and ZF receivers achieve high data rates,  which can be enhanced by increasing the number of subsymbols. In future works,  other impairments  which affect the OOB emission of SU on spectrum of PUs like as RF impairments specially nonlinear power amplifier can be investigated.
 
	%\appendices
	\begin{appendices}
	\section{}

Due to (\ref{equation100}) and (\ref{equation10}), variance of interference noise can be written by  
	 	\begin{equation}\label{equation12}
\begin{aligned}
 \sigma _{{{n}_{{{m}^{'}},{{k}^{'}}}}}^{2}=&\mathbb E[({{r}_{{{m}^{'}},{{k}^{'}}}}-{{s}_{{{m}^{'}},{{k}^{'}}}})({{r}^{*}}_{{{m}^{'}},{{k}^{'}}}-{{{s}^{*}}_{{{m}^{'}},{{k}^{'}}}})] \\ 
& =\mathbb E[{{r}_{{{m}^{'}},{{k}^{'}}}}{{r}^{*}}_{{{m}^{'}},{{k}^{'}}}]+\mathbb E[{{s}_{{{m}^{'}},{{k}^{'}}}}{{{s}^{*}}_{{{m}^{'}},{{k}^{'}}}}]\\&-2real(\mathbb E[{{r}_{{{m}^{'}},{{k}^{'}}}}{{{s}^{*}}_{{{m}^{'}},{{k}^{'}}}}]) \\ 
\end{aligned}
	 	\end{equation} 
By considering normalized prototype pulse shape ($\sum\limits_{n=0}^{MK-1}{\left|{{g}_{Tx}}[n] \right|}^{2}=1$ ), each part of (\ref{equation12}) can be calculated as

  	 	\begin{equation}\label{equation41}
\begin{aligned}
& \mathbb E[{{r}_{{{m}^{'}},{{k}^{'}}}}{{r}^{*}}_{{m}^{'},{k}^{'}}]=\frac{1}{{{\alpha }_{{{k}^{'}}}}}\sum\limits_{k=0}^{K-1}{{{\alpha }_{k}}}{{f}_{{{m}^{'}},{{k}^{'}}}}(k) \\ 
& \mathbb E[{{s}_{{{m}^{'}},{{k}^{'}}}}{{{s}^{*}}_{{{m}^{'}},{{k}^{'}}}}]=\overline{{{p}_{s}}} \\ 
& \mathbb E[{{r}_{{{m}^{'}},{{k}^{'}}}}{{{s}^{*}}_{{{m}^{'}},{{k}^{'}}}}]=\overline{{{p}_{s}}}\sum\limits_{n=0}^{MK-1}{\left| {{g}_{{{m}^{'}}}}[n] \right|^{2}}=\overline{{{p}_{s}}} \\ 
\end{aligned}
  	 	\end{equation}
According to (\ref{equation12}) and (\ref{equation41}), (\ref{equation13}) is derived.

	\section{}
According to (\ref{equation1}), the continuous form of GFDM signal by concatenating frames is expressed as

 	 	\begin{equation}\label{equation24}
x(t)=\sum\limits_{\upsilon =-\infty }^{\infty }{\sum\limits_{k=0}^{K-1}{\sum\limits_{m=0}^{M-1}{\sqrt{{{\alpha }_{k}}}{{s}_{m,k,\upsilon }}}{{g}_{T{{x}_{m}}}}(t-\upsilon {{T}_{B}}){{e}^{j2\pi \frac{(k-\frac{K-1}{2})}{{{T}_{s}}}t}}}}
 	 	\end{equation}
Where ${{g}_{T{{x}_{m}}}}(t)$ is continuous form of   ${{g}_{T{{x}_{m}}}}[n]$ with the length of $M{{T}_{s}}$ and ${{T}_{B}}$ is block duration which is equal to $M{{T}_{s}}$. To calculate its PSD, autocorrelation  of  $x(t)$ is derived as

	 	\begin{equation}\label{equation25}
        \begin{aligned}
{{R}_{xx}}(t,\tau )=&\overline{{{p}_{s}}}\sum\limits_{\upsilon =-\infty }^{\infty }\sum\limits_{k=0}^{K-1}\sum\limits_{m=0}^{M-1}{{\alpha }_{k}}{{g}_{T{{x}_{m}}}}(t-\upsilon {{T}_{B}})\\& g_{T{{x}_{m}}}^{*}(t-\tau -\upsilon {{T}_{B}}){{e}^{j2\pi \frac{(k-\frac{K-1}{2})}{{{T}_{s}}}\tau }} 
\end{aligned}
 	 	\end{equation}

Since GFDM yields cyclostationary process, ${{R}_{xx}}(t,\tau )={{R}_{xx}}(t+M{{T}_{s}},\tau )$, the average of ${{R}_{xx}}(t,\tau )$ over one period is calculated as 	 	 	 		

  	 	\begin{equation}\label{equation42}
\begin{aligned}
& {{\overline{R}}_{xx}}(\tau )=\frac{1}{M{{T}_{s}}}\int\limits_{0}^{M{{T}_{s}}}{{{R}_{xx}}(t,\tau )}dt \\ 
& =\frac{\overline{{{p}_{s}}}}{M{{T}_{s}}}(\sum\limits_{m=0}^{M-1}{{{g}_{m}}(\tau )}\otimes {{g}_{m}}(-\tau ))\sum\limits_{k=0}^{K-1}{{{\alpha }_{k}}{{e}^{j2\pi \dfrac{(k-\dfrac{K-1}{2})}{{{T}_{s}}}\tau }}} \\ 
\end{aligned}
  	 	\end{equation}

By taking the Fourier transform of (\ref{equation42}), (\ref{equation26}) is derived.
 
\end{appendices}
	\bibliographystyle{IEEE}
	\bibliography{thesis-bib}

  \end{document}

%% file: SER.tex
\begin{tikzpicture}

\begin{axis}[%
width=2.842in,
height=2.742in,
at={(0.461in,0.241in)},
scale only axis,
ticklabel style = {font=\fontsize{8}{8}\selectfont},
xmin=0,
xmax=30,
xlabel={$E_{s}/N_{0}$ (dB)},
ymode=log,
ymin=0.0001,
ymax=1,
yminorticks=true,
ylabel={SER},
axis background/.style={fill=white},
xmajorgrids,
ymajorgrids,
yminorgrids,
legend style={legend cell align=left, align=left, draw=white!15!black}
]
\addplot [color=blue, dotted, line width=1.0pt]
  table[row sep=crcr]{%
0	0.686090624999999\\
1.57894736842105	0.637365625\\
3.1578947368421	0.585584374999999\\
4.73684210526316	0.529196875\\
6.31578947368421	0.4658125\\
7.89473684210526	0.398465625\\
9.47368421052632	0.325575000000001\\
11.0526315789474	0.253846875\\
12.6315789473684	0.182509375\\
14.2105263157895	0.119734375\\
15.7894736842105	0.068246875\\
17.3684210526316	0.033096875\\
18.9473684210526	0.012690625\\
20.5263157894737	0.00330937499999999\\
21.9	0.000609375\\
23.6842105263158	2.8125e-05\\
25.2631578947368	0\\
26.8421052631579	0\\
28.4210526315789	0\\
30	0\\
};

\addplot [color=red, line width=1.0pt, draw=none, mark size=2.0pt, mark=o, mark options={solid, red}]
  table[row sep=crcr]{%
0	0.686090624999999\\
1.57894736842105	0.637365625\\
3.1578947368421	0.585584374999999\\
4.73684210526316	0.529196875\\
6.31578947368421	0.4658125\\
7.89473684210526	0.398465625\\
9.47368421052632	0.325575000000001\\
11.0526315789474	0.253846875\\
12.6315789473684	0.182509375\\
14.2105263157895	0.119734375\\
15.7894736842105	0.068246875\\
17.3684210526316	0.033096875\\
18.9473684210526	0.012690625\\
20.5263157894737	0.00330937499999999\\
21.9	0.000609375\\
23.6842105263158	2.8125e-05\\
25.2631578947368	0\\
26.8421052631579	0\\
28.4210526315789	0\\
30	0\\
};

\addplot [color=blue, dashed, line width=1.0pt]
  table[row sep=crcr]{%
0	0.692626041666667\\
1.57894736842105	0.646119791666666\\
3.15789473684211	0.5947625\\
4.73684210526316	0.537892708333334\\
6.31578947368421	0.475482291666667\\
7.89473684210526	0.407325\\
9.47368421052632	0.336971875\\
11.0526315789474	0.263289583333333\\
12.6315789473684	0.19279375\\
14.2105263157895	0.129102083333333\\
15.7894736842105	0.0751125000000001\\
17.3684210526316	0.0371052083333334\\
18.9473684210526	0.0145885416666667\\
20.5263157894737	0.00408854166666667\\
22.1052631578947	0.000731250000000005\\
23.6842105263158	7.39583333333334e-05\\
25.2631578947368	3.125e-06\\
26.8421052631579	0\\
28.4210526315789	0\\
30	0\\
};

\addplot [color=red, line width=1.0pt, draw=none, mark size=2pt, mark=triangle, mark options={solid, rotate=180, red}]
  table[row sep=crcr]{%
0	0.692626041666667\\
1.57894736842105	0.646119791666666\\
3.15789473684211	0.5947625\\
4.73684210526316	0.537892708333334\\
6.31578947368421	0.475482291666667\\
7.89473684210526	0.407325\\
9.47368421052632	0.336971875\\
11.0526315789474	0.263289583333333\\
12.6315789473684	0.19279375\\
14.2105263157895	0.129102083333333\\
15.7894736842105	0.0751125000000001\\
17.3684210526316	0.0371052083333334\\
18.9473684210526	0.0145885416666667\\
20.5263157894737	0.00408854166666667\\
22.1052631578947	0.000731250000000005\\
23.6842105263158	7.39583333333334e-05\\
25.2631578947368	3.125e-06\\
26.8421052631579	0\\
28.4210526315789	0\\
30	0\\
};

\addplot [color=blue, dashdotted, line width=1.0pt]
  table[row sep=crcr]{%
0	0.688720833333334\\
1.57894736842105	0.642894791666666\\
3.1578947368421	0.591169791666667\\
4.73684210526316	0.534367708333333\\
6.31578947368421	0.472076041666667\\
7.89473684210526	0.405551041666667\\
9.47368421052632	0.335445833333334\\
11.0526315789474	0.265325\\
12.6315789473684	0.198469791666667\\
14.2105263157895	0.137851041666667\\
15.7894736842105	0.086384375\\
17.3684210526316	0.04901875\\
18.9473684210526	0.0248083333333334\\
20.5263157894737	0.0111864583333333\\
22.1052631578947	0.00443020833333332\\
23.6842105263158	0.00158645833333334\\
25.2631578947368	0.000575000000000001\\
26.8421052631579	0.000183333333333333\\
28	7.60416666666667e-05\\
30	3.4375e-05\\
};

\addplot [color=red, line width=1.0pt, draw=none, mark size=2pt, mark=diamond, mark options={solid, red}]
  table[row sep=crcr]{%
0	0.688720833333334\\
1.57894736842105	0.642894791666666\\
3.1578947368421	0.591169791666667\\
4.73684210526316	0.534367708333333\\
6.31578947368421	0.472076041666667\\
7.89473684210526	0.405551041666667\\
9.47368421052632	0.335445833333334\\
11.0526315789474	0.265325\\
12.6315789473684	0.198469791666667\\
14.2105263157895	0.137851041666667\\
15.7894736842105	0.086384375\\
17.3684210526316	0.04901875\\
18.9473684210526	0.0248083333333334\\
20.5263157894737	0.0111864583333333\\
22.1052631578947	0.00443020833333332\\
23.6842105263158	0.00158645833333334\\
25.2631578947368	0.000575000000000001\\
26.8421052631579	0.000183333333333333\\
28	7.60416666666667e-05\\
30	3.4375e-05\\
};

\addplot [color=blue, dashed, line width=1.0pt, mark size=2.0pt, mark=o, mark options={solid, blue}]
  table[row sep=crcr]{%
0	0.6899375\\
1.57894736842105	0.6422625\\
3.1578947368421	0.591021875\\
4.73684210526316	0.53525\\
6.31578947368421	0.472984375000001\\
7.89473684210526	0.408603125\\
9.47368421052632	0.337634375\\
11.0526315789474	0.268915625\\
12.6315789473684	0.200446875\\
14.2105263157895	0.140115625\\
15.7894736842105	0.089534375\\
17.3684210526316	0.0505281250000001\\
18.9473684210526	0.026359375\\
20.5263157894737	0.012146875\\
22.1052631578947	0.00508749999999997\\
23.6842105263158	0.00199062500000001\\
25.2631578947368	0.000712500000000001\\
26.7	0.000275\\
28.1	0.00010625\\
30	5e-05\\
};

\addplot [color=red, line width=1.0pt, draw=none, mark size=2.0pt, mark=asterisk, mark options={solid, red}]
  table[row sep=crcr]{%
0	0.6899375\\
1.57894736842105	0.6422625\\
3.1578947368421	0.591021875\\
4.73684210526316	0.53525\\
6.31578947368421	0.472984375000001\\
7.89473684210526	0.408603125\\
9.47368421052632	0.337634375\\
11.0526315789474	0.268915625\\
12.6315789473684	0.200446875\\
14.2105263157895	0.140115625\\
15.7894736842105	0.089534375\\
17.3684210526316	0.0505281250000001\\
18.9473684210526	0.026359375\\
20.5263157894737	0.012146875\\
22.1052631578947	0.00508749999999997\\
23.6842105263158	0.00199062500000001\\
25.2631578947368	0.000712500000000001\\
26.7	0.000275\\
28.1	0.00010625\\
30	5e-05\\
};

\addplot [color=blue, line width=1.0pt]
  table[row sep=crcr]{%
0	0.685640625\\
1.57894736842105	0.63728125\\
3.1578947368421	0.5839375\\
4.73684210526316	0.529046875\\
6.31578947368421	0.463015625\\
7.89473684210526	0.39665625\\
9.47368421052632	0.3214375\\
11.0526315789474	0.25234375\\
12.6315789473684	0.181859375\\
14.2105263157895	0.11684375\\
15.7894736842105	0.066546875\\
17.4	0.03065625\\
18.86	0.011625\\
20.35	0.003484375\\
21.8	0.000578125\\
23.25	4.6875e-05\\
25.2631578947368	0\\
26.8421052631579	0\\
28.4210526315789	0\\
30	0\\
};

\end{axis}

\begin{axis}[%
width=0.865in,
height=0.882in,
at={(0.756in,1.247in)},
scale only axis,
ticklabel style = {font=\fontsize{8}{8}\selectfont},
xmin=20,
xmax=23,
ymode=log,
ymin=0.0001,
ymax=0.01,
yminorticks=true,
axis background/.style={fill=white},
xmajorgrids,
ymajorgrids,
yminorgrids,
legend style={legend cell align=left, align=left, draw=white!15!black}
]
\addplot [color=blue, dotted, line width=1.0pt]
  table[row sep=crcr]{%
0	0.686090624999999\\
1.57894736842105	0.637365625\\
3.1578947368421	0.585584374999999\\
4.73684210526316	0.529196875\\
6.31578947368421	0.4658125\\
7.89473684210526	0.398465625\\
9.47368421052632	0.325575000000001\\
11.0526315789474	0.253846875\\
12.6315789473684	0.182509375\\
14.2105263157895	0.119734375\\
15.7894736842105	0.068246875\\
17.3684210526316	0.033096875\\
18.9473684210526	0.012690625\\
20.5263157894737	0.00330937499999999\\
21.9	0.000609375\\
23.6842105263158	2.8125e-05\\
25.2631578947368	0\\
26.8421052631579	0\\
28.4210526315789	0\\
30	0\\
};

\addplot [color=red, line width=1.0pt, draw=none, mark size=2.0pt, mark=o, mark options={solid, red}]
  table[row sep=crcr]{%
0	0.686090624999999\\
1.57894736842105	0.637365625\\
3.1578947368421	0.585584374999999\\
4.73684210526316	0.529196875\\
6.31578947368421	0.4658125\\
7.89473684210526	0.398465625\\
9.47368421052632	0.325575000000001\\
11.0526315789474	0.253846875\\
12.6315789473684	0.182509375\\
14.2105263157895	0.119734375\\
15.7894736842105	0.068246875\\
17.3684210526316	0.033096875\\
18.9473684210526	0.012690625\\
20.5263157894737	0.00330937499999999\\
21.9	0.000609375\\
23.6842105263158	2.8125e-05\\
25.2631578947368	0\\
26.8421052631579	0\\
28.4210526315789	0\\
30	0\\
};

\addplot [color=blue, dashed, line width=1.0pt]
  table[row sep=crcr]{%
0	0.692626041666667\\
1.57894736842105	0.646119791666666\\
3.15789473684211	0.5947625\\
4.73684210526316	0.537892708333334\\
6.31578947368421	0.475482291666667\\
7.89473684210526	0.407325\\
9.47368421052632	0.336971875\\
11.0526315789474	0.263289583333333\\
12.6315789473684	0.19279375\\
14.2105263157895	0.129102083333333\\
15.7894736842105	0.0751125000000001\\
17.3684210526316	0.0371052083333334\\
18.9473684210526	0.0145885416666667\\
20.5263157894737	0.00408854166666667\\
22.1052631578947	0.000731250000000005\\
23.6842105263158	7.39583333333334e-05\\
25.2631578947368	3.125e-06\\
26.8421052631579	0\\
28.4210526315789	0\\
30	0\\
};

\addplot [color=red, line width=1.0pt, draw=none, mark size=2pt, mark=triangle, mark options={solid, rotate=180, red}]
  table[row sep=crcr]{%
0	0.692626041666667\\
1.57894736842105	0.646119791666666\\
3.15789473684211	0.5947625\\
4.73684210526316	0.537892708333334\\
6.31578947368421	0.475482291666667\\
7.89473684210526	0.407325\\
9.47368421052632	0.336971875\\
11.0526315789474	0.263289583333333\\
12.6315789473684	0.19279375\\
14.2105263157895	0.129102083333333\\
15.7894736842105	0.0751125000000001\\
17.3684210526316	0.0371052083333334\\
18.9473684210526	0.0145885416666667\\
20.5263157894737	0.00408854166666667\\
22.1052631578947	0.000731250000000005\\
23.6842105263158	7.39583333333334e-05\\
25.2631578947368	3.125e-06\\
26.8421052631579	0\\
28.4210526315789	0\\
30	0\\
};

\addplot [color=blue, dashdotted, line width=1.0pt]
  table[row sep=crcr]{%
0	0.688720833333334\\
1.57894736842105	0.642894791666666\\
3.1578947368421	0.591169791666667\\
4.73684210526316	0.534367708333333\\
6.31578947368421	0.472076041666667\\
7.89473684210526	0.405551041666667\\
9.47368421052632	0.335445833333334\\
11.0526315789474	0.265325\\
12.6315789473684	0.198469791666667\\
14.2105263157895	0.137851041666667\\
15.7894736842105	0.086384375\\
17.3684210526316	0.04901875\\
18.9473684210526	0.0248083333333334\\
20.5263157894737	0.0111864583333333\\
22.1052631578947	0.00443020833333332\\
23.6842105263158	0.00158645833333334\\
25.2631578947368	0.000575000000000001\\
26.8421052631579	0.000183333333333333\\
28	7.60416666666667e-05\\
30	3.4375e-05\\
};

\addplot [color=red, line width=1.0pt, draw=none, mark size=2pt, mark=diamond, mark options={solid, red}]
  table[row sep=crcr]{%
0	0.688720833333334\\
1.57894736842105	0.642894791666666\\
3.1578947368421	0.591169791666667\\
4.73684210526316	0.534367708333333\\
6.31578947368421	0.472076041666667\\
7.89473684210526	0.405551041666667\\
9.47368421052632	0.335445833333334\\
11.0526315789474	0.265325\\
12.6315789473684	0.198469791666667\\
14.2105263157895	0.137851041666667\\
15.7894736842105	0.086384375\\
17.3684210526316	0.04901875\\
18.9473684210526	0.0248083333333334\\
20.5263157894737	0.0111864583333333\\
22.1052631578947	0.00443020833333332\\
23.6842105263158	0.00158645833333334\\
25.2631578947368	0.000575000000000001\\
26.8421052631579	0.000183333333333333\\
28	7.60416666666667e-05\\
30	3.4375e-05\\
};

\addplot [color=blue, dashed, line width=1.0pt, mark size=2.0pt, mark=o, mark options={solid, blue}]
  table[row sep=crcr]{%
0	0.6899375\\
1.57894736842105	0.6422625\\
3.1578947368421	0.591021875\\
4.73684210526316	0.53525\\
6.31578947368421	0.472984375000001\\
7.89473684210526	0.408603125\\
9.47368421052632	0.337634375\\
11.0526315789474	0.268915625\\
12.6315789473684	0.200446875\\
14.2105263157895	0.140115625\\
15.7894736842105	0.089534375\\
17.3684210526316	0.0505281250000001\\
18.9473684210526	0.026359375\\
20.5263157894737	0.012146875\\
22.1052631578947	0.00508749999999997\\
23.6842105263158	0.00199062500000001\\
25.2631578947368	0.000712500000000001\\
26.7	0.000275\\
28.1	0.00010625\\
30	5e-05\\
};

\addplot [color=red, line width=1.0pt, draw=none, mark size=2.0pt, mark=asterisk, mark options={solid, red}]
  table[row sep=crcr]{%
0	0.6899375\\
1.57894736842105	0.6422625\\
3.1578947368421	0.591021875\\
4.73684210526316	0.53525\\
6.31578947368421	0.472984375000001\\
7.89473684210526	0.408603125\\
9.47368421052632	0.337634375\\
11.0526315789474	0.268915625\\
12.6315789473684	0.200446875\\
14.2105263157895	0.140115625\\
15.7894736842105	0.089534375\\
17.3684210526316	0.0505281250000001\\
18.9473684210526	0.026359375\\
20.5263157894737	0.012146875\\
22.1052631578947	0.00508749999999997\\
23.6842105263158	0.00199062500000001\\
25.2631578947368	0.000712500000000001\\
26.7	0.000275\\
28.1	0.00010625\\
30	5e-05\\
};

\addplot [color=blue, line width=1.0pt]
  table[row sep=crcr]{%
0	0.685640625\\
1.57894736842105	0.63728125\\
3.1578947368421	0.5839375\\
4.73684210526316	0.529046875\\
6.31578947368421	0.463015625\\
7.89473684210526	0.39665625\\
9.47368421052632	0.3214375\\
11.0526315789474	0.25234375\\
12.6315789473684	0.181859375\\
14.2105263157895	0.11684375\\
15.7894736842105	0.066546875\\
17.4	0.03065625\\
18.86	0.011625\\
20.35	0.003484375\\
21.8	0.000578125\\
23.25	4.6875e-05\\
25.2631578947368	0\\
26.8421052631579	0\\
28.4210526315789	0\\
30	0\\
};

\end{axis}

\begin{axis}[%
width=0.52in,
height=0.504in,
at={(1.86in,0.452in)},
scale only axis,
ticklabel style = {font=\fontsize{8}{8}\selectfont},
xmin=25,
xmax=27.375,
ymode=log,
ymin=0.0001,
ymax=0.001,
yminorticks=true,
axis background/.style={fill=white},
xmajorgrids,
ymajorgrids,
yminorgrids,
legend style={font=\fontsize{6.5}{6.5}\selectfont,at={(1.022,2.45)}, anchor=south west, legend cell align=left, align=left, draw=black}
]
\addplot [color=blue, dotted, line width=1.0pt]
  table[row sep=crcr]{%
0	0.686090624999999\\
1.57894736842105	0.637365625\\
3.1578947368421	0.585584374999999\\
4.73684210526316	0.529196875\\
6.31578947368421	0.4658125\\
7.89473684210526	0.398465625\\
9.47368421052632	0.325575000000001\\
11.0526315789474	0.253846875\\
12.6315789473684	0.182509375\\
14.2105263157895	0.119734375\\
15.7894736842105	0.068246875\\
17.3684210526316	0.033096875\\
18.9473684210526	0.012690625\\
20.5263157894737	0.00330937499999999\\
21.9	0.000609375\\
23.6842105263158	2.8125e-05\\
25.2631578947368	0\\
26.8421052631579	0\\
28.4210526315789	0\\
30	0\\
};
\addlegendentry{ZF-M=5(S)}

\addplot [only marks, color=red, line width=1.0pt, draw=none, mark size=2.0pt, mark=o, mark options={solid, red}]
  table[row sep=crcr]{%
0	0.686090624999999\\
1.57894736842105	0.637365625\\
3.1578947368421	0.585584374999999\\
4.73684210526316	0.529196875\\
6.31578947368421	0.4658125\\
7.89473684210526	0.398465625\\
9.47368421052632	0.325575000000001\\
11.0526315789474	0.253846875\\
12.6315789473684	0.182509375\\
14.2105263157895	0.119734375\\
15.7894736842105	0.068246875\\
17.3684210526316	0.033096875\\
18.9473684210526	0.012690625\\
20.5263157894737	0.00330937499999999\\
21.9	0.000609375\\
23.6842105263158	2.8125e-05\\
25.2631578947368	0\\
26.8421052631579	0\\
28.4210526315789	0\\
30	0\\
};
\addlegendentry{ZF-M=5(T)}

\addplot [color=blue, dashed, line width=1.0pt]
  table[row sep=crcr]{%
0	0.692626041666667\\
1.57894736842105	0.646119791666666\\
3.15789473684211	0.5947625\\
4.73684210526316	0.537892708333334\\
6.31578947368421	0.475482291666667\\
7.89473684210526	0.407325\\
9.47368421052632	0.336971875\\
11.0526315789474	0.263289583333333\\
12.6315789473684	0.19279375\\
14.2105263157895	0.129102083333333\\
15.7894736842105	0.0751125000000001\\
17.3684210526316	0.0371052083333334\\
18.9473684210526	0.0145885416666667\\
20.5263157894737	0.00408854166666667\\
22.1052631578947	0.000731250000000005\\
23.6842105263158	7.39583333333334e-05\\
25.2631578947368	3.125e-06\\
26.8421052631579	0\\
28.4210526315789	0\\
30	0\\
};
\addlegendentry{ZF-M=15(S)}

\addplot [only marks,color=red, line width=1.0pt, draw=none, mark size=2pt, mark=triangle, mark options={solid, rotate=180, red}]
  table[row sep=crcr]{%
0	0.692626041666667\\
1.57894736842105	0.646119791666666\\
3.15789473684211	0.5947625\\
4.73684210526316	0.537892708333334\\
6.31578947368421	0.475482291666667\\
7.89473684210526	0.407325\\
9.47368421052632	0.336971875\\
11.0526315789474	0.263289583333333\\
12.6315789473684	0.19279375\\
14.2105263157895	0.129102083333333\\
15.7894736842105	0.0751125000000001\\
17.3684210526316	0.0371052083333334\\
18.9473684210526	0.0145885416666667\\
20.5263157894737	0.00408854166666667\\
22.1052631578947	0.000731250000000005\\
23.6842105263158	7.39583333333334e-05\\
25.2631578947368	3.125e-06\\
26.8421052631579	0\\
28.4210526315789	0\\
30	0\\
};
\addlegendentry{ZF-M=15(T)}

\addplot [color=blue, dashdotted, line width=1.0pt]
  table[row sep=crcr]{%
0	0.688720833333334\\
1.57894736842105	0.642894791666666\\
3.1578947368421	0.591169791666667\\
4.73684210526316	0.534367708333333\\
6.31578947368421	0.472076041666667\\
7.89473684210526	0.405551041666667\\
9.47368421052632	0.335445833333334\\
11.0526315789474	0.265325\\
12.6315789473684	0.198469791666667\\
14.2105263157895	0.137851041666667\\
15.7894736842105	0.086384375\\
17.3684210526316	0.04901875\\
18.9473684210526	0.0248083333333334\\
20.5263157894737	0.0111864583333333\\
22.1052631578947	0.00443020833333332\\
23.6842105263158	0.00158645833333334\\
25.2631578947368	0.000575000000000001\\
26.8421052631579	0.000183333333333333\\
28	7.60416666666667e-05\\
30	3.4375e-05\\
};
\addlegendentry{MF-M=5(S)}

\addplot [only marks,color=red, line width=1.0pt, draw=none, mark size=2pt, mark=diamond, mark options={solid, red}]
  table[row sep=crcr]{%
0	0.688720833333334\\
1.57894736842105	0.642894791666666\\
3.1578947368421	0.591169791666667\\
4.73684210526316	0.534367708333333\\
6.31578947368421	0.472076041666667\\
7.89473684210526	0.405551041666667\\
9.47368421052632	0.335445833333334\\
11.0526315789474	0.265325\\
12.6315789473684	0.198469791666667\\
14.2105263157895	0.137851041666667\\
15.7894736842105	0.086384375\\
17.3684210526316	0.04901875\\
18.9473684210526	0.0248083333333334\\
20.5263157894737	0.0111864583333333\\
22.1052631578947	0.00443020833333332\\
23.6842105263158	0.00158645833333334\\
25.2631578947368	0.000575000000000001\\
26.8421052631579	0.000183333333333333\\
28	7.60416666666667e-05\\
30	3.4375e-05\\
};
\addlegendentry{MF-M=5(T)}

\addplot [color=blue, dashed, line width=1.0pt, mark size=2.0pt, mark=o, mark options={solid, blue}]
  table[row sep=crcr]{%
0	0.6899375\\
1.57894736842105	0.6422625\\
3.1578947368421	0.591021875\\
4.73684210526316	0.53525\\
6.31578947368421	0.472984375000001\\
7.89473684210526	0.408603125\\
9.47368421052632	0.337634375\\
11.0526315789474	0.268915625\\
12.6315789473684	0.200446875\\
14.2105263157895	0.140115625\\
15.7894736842105	0.089534375\\
17.3684210526316	0.0505281250000001\\
18.9473684210526	0.026359375\\
20.5263157894737	0.012146875\\
22.1052631578947	0.00508749999999997\\
23.6842105263158	0.00199062500000001\\
25.2631578947368	0.000712500000000001\\
26.7	0.000275\\
28.1	0.00010625\\
30	5e-05\\
};
\addlegendentry{MF-M=15(S)}

\addplot [only marks,color=red, line width=1.0pt, draw=none, mark size=2.0pt, mark=asterisk, mark options={solid, red}]
  table[row sep=crcr]{%
0	0.6899375\\
1.57894736842105	0.6422625\\
3.1578947368421	0.591021875\\
4.73684210526316	0.53525\\
6.31578947368421	0.472984375000001\\
7.89473684210526	0.408603125\\
9.47368421052632	0.337634375\\
11.0526315789474	0.268915625\\
12.6315789473684	0.200446875\\
14.2105263157895	0.140115625\\
15.7894736842105	0.089534375\\
17.3684210526316	0.0505281250000001\\
18.9473684210526	0.026359375\\
20.5263157894737	0.012146875\\
22.1052631578947	0.00508749999999997\\
23.6842105263158	0.00199062500000001\\
25.2631578947368	0.000712500000000001\\
26.7	0.000275\\
28.1	0.00010625\\
30	5e-05\\
};
\addlegendentry{MF-M=15(T)}

\addplot [color=blue, line width=1.0pt]
  table[row sep=crcr]{%
0	0.685640625\\
1.57894736842105	0.63728125\\
3.1578947368421	0.5839375\\
4.73684210526316	0.529046875\\
6.31578947368421	0.463015625\\
7.89473684210526	0.39665625\\
9.47368421052632	0.3214375\\
11.0526315789474	0.25234375\\
12.6315789473684	0.181859375\\
14.2105263157895	0.11684375\\
15.7894736842105	0.066546875\\
17.4	0.03065625\\
18.86	0.011625\\
20.35	0.003484375\\
21.8	0.000578125\\
23.25	4.6875e-05\\
25.2631578947368	0\\
26.8421052631579	0\\
28.4210526315789	0\\
30	0\\
};
\addlegendentry{OFDM}

\end{axis}

\begin{axis}[%
width=3.667in,
height=2.99in,
at={(0in,0in)},
scale only axis,
xmin=0,
xmax=1,
ymin=0,
ymax=1,
axis line style={draw=none},
ticks=none,
axis x line*=bottom,
axis y line*=left,
legend style={legend cell align=left, align=left, draw=white!15!black}
]
\draw [black] (axis cs:0.781933,0.271569) ellipse [x radius=0.0294708, y radius=0.0303922];
\draw [black] (axis cs:0.651085,0.434314) ellipse [x radius=0.0331257, y radius=0.0303922];
\draw[thick,->] (0.755933,0.251569) -- (0.651933,0.19569) ;
\draw[thick,->] (0.621085,0.434314) -- (0.441933,0.49569) ;
\end{axis}
\end{tikzpicture}%

%% file: Qn.tex
\definecolor{mycolor1}{rgb}{0.84706,0.16078,0.00000}%
\begin{tikzpicture}

\begin{axis}[%
width=2.842in,
height=2.742in,
at={(0.482in,0.289in)},
scale only axis,
ticklabel style = {font=\fontsize{8}{8}\selectfont},
xmin=0,
xmax=0.5,
xlabel={$Q_n$},
ymin=0,
ymax=4,
ylabel={Capacity(bit/Hz)},
axis background/.style={fill=white},
xmajorgrids,
ymajorgrids,
legend style={font=\fontsize{7}{7}\selectfont, at={(0.31,0.044)}, anchor=south west, legend cell align=left, align=left, draw=black}
]
\addplot [color=blue, line width=2.0pt]
  table[row sep=crcr]{%
0	-2.62185947230243e-14\\
0.01	1.59937878560856\\
0.02	2.32177391235864\\
0.03	2.77228612639904\\
0.04	3.07233591349001\\
0.05	3.27218288464029\\
0.06	3.40690332871372\\
0.07	3.4964324154986\\
0.08	3.5548528099742\\
0.09	3.59208926283262\\
0.1	3.61548296488653\\
0.11	3.62926143707623\\
0.12	3.63719228918535\\
0.13	3.64107167297468\\
0.14	3.64266437984323\\
0.15	3.64281122890602\\
0.16	3.64212638625915\\
0.17	3.64067118127773\\
0.18	3.639125795244\\
0.19	3.6375451496103\\
0.2	3.63555685494572\\
0.21	3.63319440961726\\
0.22	3.63075087681677\\
0.23	3.62854174832843\\
0.24	3.62661729971197\\
0.25	3.62474451936694\\
0.26	3.62295692479922\\
0.27	3.62135806892192\\
0.28	3.61985268899537\\
0.29	3.61840763398981\\
0.3	3.61695353133074\\
0.31	3.61555454789742\\
0.32	3.61419097268103\\
0.33	3.61289189746092\\
0.34	3.61165072503208\\
0.35	3.61044813835156\\
0.36	3.60921383117477\\
0.37	3.60793191015482\\
0.38	3.60659930438073\\
0.39	3.60521487153652\\
0.4	3.60384375191632\\
0.41	3.60241743381351\\
0.42	3.60097704350287\\
0.43	3.59952730119474\\
0.44	3.59800412771883\\
0.45	3.59639910178594\\
0.46	3.59492546044861\\
0.47	3.59340385127016\\
0.48	3.59197437838768\\
0.49	3.59068265259399\\
0.5	3.58934950114668\\
};
\addlegendentry{$\text{Q}_{\text{r}}\text{=Q}_{\text{l}}\text{=5 dBm}$}

\addplot [color=mycolor1, dashed, line width=2.0pt]
  table[row sep=crcr]{%
0	7.79662523608745e-14\\
0.01	1.56595957763297\\
0.02	2.10540032650551\\
0.03	2.3071239299737\\
0.04	2.37747492514308\\
0.05	2.40165285014963\\
0.06	2.40857232670632\\
0.07	2.40876261844657\\
0.08	2.40697677424297\\
0.09	2.40454470445555\\
0.1	2.40279417871527\\
0.11	2.40092501634071\\
0.12	2.39935115449665\\
0.13	2.39782026559873\\
0.14	2.39618692296291\\
0.15	2.39456777587794\\
0.16	2.39296132938651\\
0.17	2.39133677034455\\
0.18	2.38970803456643\\
0.19	2.38803437222362\\
0.2	2.3863504653056\\
0.21	2.38456837905924\\
0.22	2.38270490933612\\
0.23	2.38078701555788\\
0.24	2.3789078315241\\
0.25	2.37690385367202\\
0.26	2.37481257612706\\
0.27	2.37271705553649\\
0.28	2.37081628089066\\
0.29	2.3689411111328\\
0.3	2.36724942470783\\
0.31	2.36552544267239\\
0.32	2.36373331119805\\
0.33	2.36199868338559\\
0.34	2.36039902859328\\
0.35	2.35884147049788\\
0.36	2.35724285639973\\
0.37	2.35557517589855\\
0.38	2.35386601982334\\
0.39	2.35218687809031\\
0.4	2.35050345906129\\
0.41	2.34875574566133\\
0.42	2.34693064890089\\
0.43	2.34510249873481\\
0.44	2.34319883207359\\
0.45	2.34127921027536\\
0.46	2.3394909822258\\
0.47	2.33773576266413\\
0.48	2.33594102312242\\
0.49	2.33409479074544\\
0.5	2.33230419267887\\
};
\addlegendentry{$\text{Q}_{\text{r}}\text{=Q}_{\text{l}}\text{=0 dBm}$}

\addplot [color=black, dotted, line width=2.0pt]
  table[row sep=crcr]{%
0	-7.82574340545158e-15\\
0.000408163265306122	0.122948645217031\\
0.000816326530612245	0.233787989457059\\
0.00122448979591837	0.334794813051774\\
0.00163265306122449	0.427537054784335\\
0.00204081632653061	0.512563408001763\\
0.00244897959183673	0.591215036954707\\
0.00285714285714286	0.664056244034836\\
0.00326530612244898	0.731296856142372\\
0.0036734693877551	0.79279786421037\\
0.00408163265306122	0.848345112702753\\
0.00448979591836735	0.898597756106091\\
0.00489795918367347	0.944276127838934\\
0.00530612244897959	0.986048087376206\\
0.00571428571428572	1.02424636049714\\
0.00612244897959184	1.05891499683931\\
0.00653061224489796	1.09032719292319\\
0.00693877551020408	1.11891321575657\\
0.0073469387755102	1.14517932849366\\
0.00775510204081633	1.16945527229333\\
0.00816326530612245	1.19176928011188\\
0.00857142857142857	1.21210271605983\\
0.00897959183673469	1.23062216176198\\
0.00938775510204082	1.24738953824606\\
0.00979591836734694	1.26249827334136\\
0.0102040816326531	1.27612383989331\\
0.0106122448979592	1.28850807375318\\
0.0110204081632653	1.29972718036509\\
0.0114285714285714	1.30985364589047\\
0.0118367346938776	1.31905638904298\\
0.0122448979591837	1.32739508729541\\
0.0126530612244898	1.3349889381259\\
0.0130612244897959	1.34186512959383\\
0.013469387755102	1.3481420010236\\
0.0138775510204082	1.35390445995225\\
0.0142857142857143	1.35918663922589\\
0.0146938775510204	1.36403786500268\\
0.0151020408163265	1.36848075430074\\
0.0155102040816327	1.37257635888553\\
0.0159183673469388	1.3763426087717\\
0.0163265306122449	1.37982988871328\\
0.016734693877551	1.3830566079139\\
0.0171428571428571	1.38603224426451\\
0.0175510204081633	1.38877824516889\\
0.0179591836734694	1.39132909925223\\
0.0183673469387755	1.39367727806493\\
0.0187755102040816	1.3958307516909\\
0.0191836734693878	1.39781066233378\\
0.0195918367346939	1.39961399223962\\
0.02	1.40126524085499\\
0.03	1.41433217669128\\
0.0822222222222222	1.40729655166071\\
0.134444444444444	1.399610046706\\
0.186666666666667	1.39146929930809\\
0.238888888888889	1.38341285610809\\
0.291111111111111	1.37566781218924\\
0.343333333333333	1.36849783229349\\
0.395555555555556	1.36136255365918\\
0.447777777777778	1.35463535777851\\
0.5	1.34763204483863\\
};
\addlegendentry{$\text{Q}_{\text{r}}\text{=Q}_{\text{l}}\text{=-5 dBm}$}

\end{axis}
\end{tikzpicture}%

%% file: MF.tex
\definecolor{mycolor1}{rgb}{0.84706,0.16078,0.00000}%
\begin{tikzpicture}

\begin{axis}[%
width=2.842in,
height=2.742in,
at={(0.466in,0.384in)},
scale only axis,
ticklabel style = {font=\fontsize{8}{8}\selectfont},
xmin=-20,
xmax=35,
xlabel={$Q_{r}=Q_{l}\text{ (dBm)}$},
ymin=-0.5,
ymax=2.5,
ylabel={Capacity (bit/s/Hz)},
axis background/.style={fill=white},
xmajorgrids,
ymajorgrids,
legend style={{font=\fontsize{7}{7}\selectfont}, at={(0.055,0.585)}, anchor=south west, legend cell align=left, align=left, draw=black}
]
\addplot [color=blue, line width=1.0pt, mark size=2.0pt, mark=asterisk, mark options={solid, blue}]
  table[row sep=crcr]{%
-20	0.046098276792071\\
-15	0.09\\
-10	0.16\\
-5	0.28484481570012\\
0	0.484683746119304\\
5	0.730923479133413\\
10	0.981745456855063\\
15	1.19224061480669\\
20	1.33575558198087\\
25	1.3667375\\
30	1.368\\
35	1.368\\
};
\addlegendentry{M=5, Non-uniform}

\addplot [color=mycolor1, line width=1.0pt, mark size=2pt, mark=triangle, mark options={solid, rotate=90, mycolor1}]
  table[row sep=crcr]{%
-20	0.00385922899812644\\
-15	0.0118749166143163\\
-10	0.0349016365518265\\
-5	0.0935094193924149\\
0	0.21721611807736\\
5	0.424500910292826\\
10	0.696765722882623\\
15	0.97872030568546\\
20	1.21120687013419\\
25	1.3449515625\\
30	1.3679578125\\
35	1.3679578125\\
};
\addlegendentry{M=5, Uniform}

\addplot [color=blue, line width=1.0pt, mark size=2pt, mark=square, mark options={solid, blue}]
  table[row sep=crcr]{%
-20	0.124376599452283\\
-15	0.22\\
-10	0.320847381048913\\
-5	0.463189579244767\\
0	0.688958737830499\\
5	0.91272369107268\\
10	1.09672137802019\\
15	1.22022503744034\\
20	1.25272772788082\\
25	1.25537065311953\\
30	1.25539284520388\\
35	1.25539284520388\\
};
\addlegendentry{M=15, Non-uniform}

\addplot [color=mycolor1, line width=1pt, mark size=2pt, mark=diamond, mark options={solid, mycolor1}]
  table[row sep=crcr]{%
-20	0.00891732456982883\\
-15	0.0264866186256345\\
-10	0.0724077274207891\\
-5	0.173240148806455\\
0	0.350143299789671\\
5	0.592546676278139\\
10	0.851819392503698\\
15	1.07088503071549\\
20	1.21078730523794\\
25	1.25425301657297\\
30	1.25526520245071\\
35	1.25526520245071\\
};
\addlegendentry{M=15, Uniform}

\addplot [color=black, line width=1pt, mark size=2pt, mark=+, mark options={solid, black}]
  table[row sep=crcr]{%
-20	0.00661801599591893\\
-15	0.0179025358978844\\
-10	0.0445879392274459\\
-5	0.101240870509954\\
0	0.209014196077422\\
5	0.3903003184673\\
10	0.657739789900975\\
15	1.00675160698421\\
20	1.41997319705638\\
25	1.860728125\\
30	2.0243203125\\
35	2.02925625\\
};
\addlegendentry{OFDM, Non-uniform}

\addplot [color=black, line width=1.0pt, mark size=2.0pt, mark=o, mark options={solid, black}]
  table[row sep=crcr]{%
-20	0.000828516127823575\\
-15	0.00260582439776218\\
-10	0.00810561048215802\\
-5	0.0244605787488394\\
0	0.0688971333046038\\
5	0.172117440482423\\
10	0.367331909409599\\
15	0.666712855717531\\
20	1.05333095809582\\
25	1.4970015625\\
30	1.9090953125\\
35	2.0264625\\
};
\addlegendentry{OFDM, Uniform}

\end{axis}
\end{tikzpicture}%

%% file: ZF.tex
\definecolor{mycolor1}{rgb}{0.84706,0.16078,0.00000}%
\begin{tikzpicture}

\begin{axis}[%
width=2.842in,
height=2.742in,
at={(0.477in,0.382in)},
scale only axis,
xmin=-20,
xmax=35,
ticklabel style = {font=\fontsize{8}{8}\selectfont},
xlabel={$Q_{r}=Q_{l}\text{ (dBm)}$},
ymin=-0.5,
ymax=2.5,
ylabel={Capacity(bit/s/Hz)},
axis background/.style={fill=white},
xmajorgrids,
ymajorgrids,
legend style={{font=\fontsize{7}{7}\selectfont}, at={(0.04,0.585)}, anchor=south west, legend cell align=left, align=left, draw=black}
]
\addplot [color=blue, line width=1.0pt, mark size=2.0pt, mark=asterisk, mark options={solid, blue}]
  table[row sep=crcr]{%
-20	0.0294806435727591\\
-15	0.0698738182001524\\
-10	0.150784926881424\\
-5	0.294887735489516\\
0	0.520601616295078\\
5	0.831255211406813\\
10	1.2144369702819\\
15	1.64990450345058\\
20	2.00622188662972\\
25	2.0489828125\\
30	2.05003125\\
35	2.05003125\\
};
\addlegendentry{M=5, Non-uniform}

\addplot [color=mycolor1, line width=1.0pt, mark size=2pt, mark=triangle, mark options={solid, rotate=270, mycolor1}]
  table[row sep=crcr]{%
-20	0.00370084763745652\\
-15	0.0114152539229377\\
-10	0.033753918313878\\
-5	0.0915613539784745\\
0	0.217337181796861\\
5	0.439887616265694\\
10	0.76327384303967\\
15	1.16593766128193\\
20	1.61558268866417\\
25	1.973896875\\
30	2.047703125\\
35	2.047703125\\
};
\addlegendentry{M=5, Uniform}

\addplot [color=blue, line width=1pt, mark size=2pt, mark=square, mark options={solid, blue}]
  table[row sep=crcr]{%
-20	0.0604998162349537\\
-15	0.132454303176917\\
-10	0.263162469358883\\
-5	0.472459821479811\\
0	0.766908063244136\\
5	1.13692580973253\\
10	1.56326602909834\\
15	1.92867869430711\\
20	1.97984750084852\\
25	1.982234375\\
30	1.98225\\
35	1.98225\\
};
\addlegendentry{M=15, Non-uniform}

\addplot [color=mycolor1, line width=1pt, mark size=2pt, mark=diamond, mark options={solid, mycolor1}]
  table[row sep=crcr]{%
-20	0.007258985187917\\
-15	0.021850071719281\\
-10	0.061372535577966\\
-5	0.153506542488697\\
0	0.330547679900342\\
5	0.609057964586332\\
10	0.978236543035489\\
15	1.41009097469222\\
20	1.81143949600229\\
25	1.975465625\\
30	1.982\\
35	1.982\\
};
\addlegendentry{M=15, Uniform}

\addplot [color=black, line width=1.0pt, mark size=2.0pt, mark=+, mark options={solid, black}]
  table[row sep=crcr]{%
-20	0.00661801599591893\\
-15	0.0179025358978844\\
-10	0.0445879392274459\\
-5	0.101240870509954\\
0	0.209014196077422\\
5	0.3903003184673\\
10	0.657739789900975\\
15	1.00675160698421\\
20	1.41997319705638\\
25	1.860728125\\
30	2.0243203125\\
35	2.02925625\\
};
\addlegendentry{OFDM, Non-uniform}

\addplot [color=black, line width=1.0pt, mark=o, mark options={solid, black}]
  table[row sep=crcr]{%
-20	0.000828516127823575\\
-15	0.00260582439776218\\
-10	0.00810561048215802\\
-5	0.0244605787488394\\
0	0.0688971333046038\\
5	0.172117440482423\\
10	0.367331909409599\\
15	0.666712855717531\\
20	1.05333095809582\\
25	1.4970015625\\
30	1.9090953125\\
35	2.0264625\\
};
\addlegendentry{OFDM, Uniform}

\end{axis}

\begin{axis}[%
width=0.852in,
height=0.74in,
at={(2.388in,0.59in)},
ticklabel style = {font=\fontsize{8}{8}\selectfont},
scale only axis,
xmin=22,
xmax=35,
ymin=1.8,
ymax=2.1,
axis background/.style={fill=white},
xmajorgrids,
ymajorgrids,
legend style={legend cell align=left, align=left, draw=white!15!black}
]
\addplot [color=mycolor1, line width=1.0pt, mark size=2pt, mark=triangle, mark options={solid, rotate=270, mycolor1}]
  table[row sep=crcr]{%
-20	0.00370084763745652\\
-15	0.0114152539229377\\
-10	0.033753918313878\\
-5	0.0915613539784745\\
0	0.217337181796861\\
5	0.439887616265694\\
10	0.76327384303967\\
15	1.16593766128193\\
20	1.61558268866417\\
25	1.973896875\\
30	2.047703125\\
35	2.047703125\\
};

\addplot [color=blue, line width=1.0pt, mark size=2.0pt, mark=asterisk, mark options={solid, blue}]
  table[row sep=crcr]{%
-20	0.0294806435727591\\
-15	0.0698738182001524\\
-10	0.150784926881424\\
5	0.294887735489516\\
0	0.520601616295078\\
5	0.831255211406813\\
10	1.2144369702819\\
15	1.64990450345058\\
20	2.00622188662972\\
25	2.0489828125\\
30	2.05003125\\
35	2.05003125\\
};

\addplot [color=blue, line width=1.0pt, mark size=2pt, mark=square, mark options={solid, blue}]
  table[row sep=crcr]{%
-20	0.0604998162349537\\
-15	0.132454303176917\\
-10	0.263162469358883\\
-5	0.472459821479811\\
0	0.766908063244136\\
5	1.13692580973253\\
10	1.56326602909834\\
15	1.92867869430711\\
20	1.97984750084852\\
25	1.982234375\\
30	1.98225\\
35	1.98225\\
};

\addplot [color=mycolor1, line width=1.0pt, mark size=2pt, mark=diamond, mark options={solid, mycolor1}]
  table[row sep=crcr]{%
-20	0.007258985187917\\
-15	0.021850071719281\\
-10	0.061372535577966\\
-5	0.153506542488697\\
0	0.330547679900342\\
5	0.609057964586332\\
10	0.978236543035489\\
15	1.41009097469222\\
20	1.81143949600229\\
25	1.975465625\\
30	1.982\\
35	1.982\\
};

\addplot [color=black, line width=1.0pt, mark size=2.0pt, mark=+, mark options={solid, black}]
  table[row sep=crcr]{%
-20	0.00661801599591893\\
-15	0.0179025358978844\\
-10	0.0445879392274459\\
-5	0.101240870509954\\
0	0.209014196077422\\
5	0.3903003184673\\
10	0.657739789900975\\
15	1.00675160698421\\
20	1.41997319705638\\
25	1.860728125\\
30	2.0243203125\\
35	2.02925625\\
};

\addplot [color=black, line width=1.0pt, mark=o, mark options={solid, black}]
  table[row sep=crcr]{%
-20	0.000828516127823575\\
-15	0.00260582439776218\\
-10	0.00810561048215802\\
-5	0.0244605787488394\\
0	0.0688971333046038\\
5	0.172117440482423\\
10	0.367331909409599\\
15	0.666712855717531\\
20	1.05333095809582\\
25	1.4970015625\\
30	1.9090953125\\
35	2.0264625\\
};

\end{axis}

\begin{axis}[%
width=3.667in,
height=2.99in,
at={(0in,0in)},
scale only axis,
xmin=0,
xmax=1,
ymin=0,
ymax=1,
axis line style={draw=none},
ticks=none,
axis x line*=bottom,
axis y line*=left,
legend style={legend cell align=left, align=left, draw=white!15!black}
]
\draw [black] (axis cs:0.825792,0.899564) ellipse [x radius=0.0718684, y radius=0.0675381];
\draw[thick,->] (0.825792,0.829564) -- (0.80342,0.458429) ;
\end{axis}
\end{tikzpicture}%

%% file: Joft.tex
\definecolor{mycolor1}{rgb}{0.84706,0.16078,0.00000}%
\begin{tikzpicture}

\begin{axis}[%
width=2.842in,
height=2.742in,
at={(0.3in,0.2in)},
scale only axis,
ticklabel style = {font=\fontsize{8}{8}\selectfont},
xmin=-20,
xmax=35,
xlabel={$Q_{r}=Q_{l}\text{ (dBm)}$},
ymin=-.1,
ymax=2.2,
ylabel={Capacity(bit/s/Hz)},
axis background/.style={fill=white},
xmajorgrids,
ymajorgrids,
legend style={{font=\fontsize{7}{7}\selectfont},at={(0.083,0.533)}, anchor=south west, legend cell align=left, align=left, draw=black}
]
\addplot [color=blue, line width=1.0pt, mark size=2.0pt, mark=asterisk, mark options={solid, blue}]
  table[row sep=crcr]{%
-20	0.0294806435727591\\
-15	0.0698738182001524\\
-10	0.150784926881424\\
-5	0.294887735489516\\
0	0.520601616295078\\
5	0.831255211406813\\
10	1.2144369702819\\
15	1.64990450345058\\
20	2.00622188662972\\
25	2.0489828125\\
30	2.05003125\\
35	2.05003125\\
};
\addlegendentry{ZF,Non-uniform}

\addplot [color=mycolor1, line width=1.0pt, mark size=2pt, mark=triangle, mark options={solid, rotate=270, mycolor1}]
  table[row sep=crcr]{%
-20	0.00370084763745652\\
-15	0.0114152539229377\\
-10	0.033753918313878\\
-5	0.0915613539784745\\
0	0.217337181796861\\
5	0.439887616265694\\
10	0.76327384303967\\
15	1.16593766128193\\
20	1.61558268866417\\
25	1.973896875\\
30	2.047703125\\
35	2.047703125\\
};
\addlegendentry{ZF, Uniform}

\addplot [color=blue, line width=1pt, mark size=2pt, mark=square, mark options={solid, blue}]
  table[row sep=crcr]{%
-20	0.046098276792071\\
-15	0.09\\
-10	0.16\\
-5	0.28484481570012\\
0	0.484683746119304\\
5	0.730923479133413\\
10	0.981745456855063\\
15	1.19224061480669\\
20	1.33575558198087\\
25	1.3667375\\
30	1.368\\
35	1.368\\
};
\addlegendentry{MF, Non-uniform}

\addplot [color=mycolor1, line width=1pt, mark size=2pt, mark=o, mark options={solid, mycolor1}]
  table[row sep=crcr]{%
-20	0.00385922899812644\\
-15	0.0118749166143163\\
-10	0.0349016365518265\\
-5	0.0935094193924149\\
0	0.21721611807736\\
5	0.424500910292826\\
10	0.696765722882623\\
15	0.97872030568546\\
20	1.21120687013419\\
25	1.3449515625\\
30	1.3679578125\\
35	1.3679578125\\
};
\addlegendentry{MF, Uniform}

\end{axis}

\begin{axis}[%
width=0.865in,
height=0.697in,
at={(2.206in,0.574in)},
ticklabel style = {font=\fontsize{8}{8}\selectfont},
scale only axis,
xmin=-20,
xmax=0,
ymin=0.01,
ymax=0.2,
axis background/.style={fill=white},
xmajorgrids,
ymajorgrids,
legend style={legend cell align=left, align=left, draw=white!15!black}
]
\addplot [color=blue, line width=1pt, mark size=2pt, mark=asterisk, mark options={solid, blue}]
  table[row sep=crcr]{%
-20	0.0294806435727591\\
-15	0.0698738182001524\\
-10	0.150784926881424\\
-5	0.294887735489516\\
0	0.520601616295078\\
5	0.831255211406813\\
10	1.2144369702819\\
15	1.64990450345058\\
20	2.00622188662972\\
25	2.0489828125\\
30	2.05003125\\
35	2.05003125\\
};

\addplot [color=mycolor1, line width=1pt, mark size=2pt, mark=triangle, mark options={solid, rotate=270, mycolor1}]
  table[row sep=crcr]{%
-20	0.00370084763745652\\
-15	0.0114152539229377\\
-10	0.033753918313878\\
-5	0.0915613539784745\\
0	0.217337181796861\\
5	0.439887616265694\\
10	0.76327384303967\\
15	1.16593766128193\\
20	1.61558268866417\\
25	1.973896875\\
30	2.047703125\\
35	2.047703125\\
};

\addplot [color=blue, line width=1pt, mark size=2pt, mark=square, mark options={solid, blue}]
  table[row sep=crcr]{%
-20	0.046098276792071\\
-15	0.09\\
-10	0.16\\
-5	0.28484481570012\\
0	0.484683746119304\\
5	0.730923479133413\\
10	0.981745456855063\\
15	1.19224061480669\\
20	1.33575558198087\\
25	1.3667375\\
30	1.368\\
35	1.368\\
};

\addplot [color=mycolor1, line width=1pt, mark size=2pt, mark=o, mark options={solid, mycolor1}]
  table[row sep=crcr]{%
-20	0.00385922899812644\\
-15	0.0118749166143163\\
-10	0.0349016365518265\\
-5	0.0935094193924149\\
0	0.21721611807736\\
5	0.424500910292826\\
10	0.696765722882623\\
15	0.97872030568546\\
20	1.21120687013419\\
25	1.3449515625\\
30	1.3679578125\\
35	1.3679578125\\
};

\end{axis}

\begin{axis}[%
width=3.667in,
height=2.99in,
at={(0in,0in)},
scale only axis,
xmin=0,
xmax=1,
ymin=0,
ymax=1,
axis line style={draw=none},
ticks=none,
axis x line*=bottom,
axis y line*=left,
legend style={legend cell align=left, align=left, draw=white!15!black}
]
\draw [black] (axis cs:0.215985,0.142667) ellipse [x radius=0.117518, y radius=0.0706667];
\draw[thick,->] (0.335985,0.142667) -- (0.59342,0.208429) ;
\end{axis}
\end{tikzpicture}%

%% file: power.tex
\begin{tikzpicture}

\begin{axis}[%
width=2.842in,
height=2.742in,
at={(0.637in,0.659in)},
scale only axis,
ticklabel style = {font=\fontsize{8}{8}\selectfont},
xmin=-20,
xmax=35,
xlabel={$Q_{r}=Q_{l}\text{ (dBm)}$},
ymin=10,
ymax=60,
ylabel={Total allocated power (dBm)},
axis background/.style={fill=white},
xmajorgrids,
ymajorgrids,
legend style={{font=\fontsize{7}{7}\selectfont},at={(0.444,0.183)}, anchor=south west, legend cell align=left, align=left, draw=black}
]
\addplot [color=blue, line width=1.0pt, mark size=2pt, mark=square, mark options={solid, blue}]
  table[row sep=crcr]{%
-20	18.2215\\
-15	23.17\\
-10	28.05\\
-5	32.86422165090393\\
0	37.65849453317648\\
5	42.4605385484731\\
10	47.2912404911854\\
15	52.1565657381121\\
20	54.9704955710222\\
25	55.000000466711\\
30	55.000000466711\\
35	55.000000466711\\
};
\addlegendentry{GFDM-ZF,M=5}

\addplot [color=blue, line width=1.0pt, mark size=2pt, mark=triangle, mark options={solid, rotate=270, blue}]
  table[row sep=crcr]{%
-20	23\\
-15	27.83\\
-10	32.65748515133008\\
-5	37.46078973320109\\
0	42.2667841383149\\
5	47.0917514600011\\
10	51.9540594715707\\
15	54.9556314692011\\
20	54.9999999999931\\
25	55.000000466711\\
30	55.000000466711\\
35	55.000000466711\\
};
\addlegendentry{GFDM-ZF,M=15}

\addplot [color=red, line width=1.0pt, mark size=2pt, mark=triangle, mark options={solid, red}]
  table[row sep=crcr]{%
-20	20.9\\
-15	24.2\\
-10	28.1\\
-5	32.88095746641745\\
0	37.68391061248218\\
5	42.3964813814371\\
10	46.9869792954889\\
15	51.0698452979104\\
20	54.6909722979063\\
25	54.9844936796549\\
30	54.983105537896\\
35	54.9942161458674\\
};
\addlegendentry{GFDM-MF,M=5}

\addplot [color=red, line width=1.0pt, mark size=2.0pt, mark=asterisk, mark options={solid, red}]
  table[row sep=crcr]{%
-20	27.53\\
-15	30.5\\
-10	33.94349004992448\\
-5	37.47701897316722\\
0	42.1236664847812\\
5	46.7677851514104\\
10	50.9684017195829\\
15	54.6480270773012\\
20	54.9732734181363\\
25	54.9923520858051\\
30	54.9927391585478\\
35	54.9927389406386\\
};
\addlegendentry{GFDM-MF,M=15}

\addplot [color=black, line width=1.0pt, mark size=2.0pt, mark=o, mark options={solid, black}]
  table[row sep=crcr]{%
-20	10.86\\
-15	15.85\\
-10	20.82\\
-5	25.73\\
0	30.56759197087919\\
5	35.36634859387773\\
10	40.161779101951\\
15	44.9713531251578\\
20	49.8218049733936\\
25	54.2662364288549\\
30	55.000000466711\\
35	55.000000466711\\
};
\addlegendentry{OFDM}

\end{axis}
\end{tikzpicture}%

%% file: tad.tex
\definecolor{mycolor1}{rgb}{0.84706,0.16078,0.00000}%
\begin{tikzpicture}

\begin{axis}[%
width=2.842in,
height=2.742in,
at={(0.471in,0.404in)},
scale only axis,
ticklabel style = {font=\fontsize{8}{8}\selectfont},
xmin=-20,
xmax=35,
xlabel={$Q_{r}=Q_{l}\text{ (dBm)}$},
ymin=-25,
ymax=40,
ylabel={Interference power (dBm)},
axis background/.style={fill=white},
xmajorgrids,
ymajorgrids,
legend style={{font=\fontsize{7}{7}\selectfont}, at={(0.125,0.628)}, anchor=south west, legend cell align=left, align=left, draw=black}
]
\addplot [color=blue, line width=1.0pt, mark size=2.0pt, mark=asterisk, mark options={solid, blue}]
  table[row sep=crcr]{%
-20	-20\\
-15	-15\\
-10	-9.9975722257885\\
-5	-5.0171191191976\\
0	-0.0147523188261\\
5	5.0152073879969\\
10	10.0130729554371\\
15	15.0349880285853\\
20	20.1242374750689\\
25	24.56197666838345\\
30	25.54084586048877\\
35	25.54084586048877\\
};
\addlegendentry{GFDM-ZF}

\addplot [color=mycolor1, line width=1.0pt, draw=none, mark size=2pt, mark=square, mark options={solid, mycolor1}]
  table[row sep=crcr]{%
-20	-20.029003636402\\
-15	-15.0126338515502\\
-10	-10.0149649280958\\
-5	-5.0181194758353\\
0	-0.0149034136017\\
5	5.0154087265225\\
10	10.0134347336741\\
15	15.0120183677208\\
20	20.1195244978119\\
25	24.59159450187668\\
30	25.50984683652214\\
35	25.50984683652214\\
};
\addlegendentry{GFDM-MF}

\addplot [color=black, line width=1.0pt, mark size=2.0pt, mark=o, mark options={solid, black}]
  table[row sep=crcr]{%
-20	-20.0863324023648\\
-15	-15.0450184673934\\
-10	-10.0160135660912\\
-5	-5.0141523431882\\
0	-0.016876820905\\
5	5.0144610825608\\
10	10.0152748909561\\
15	15.0144951803484\\
20	20.0123562399729\\
25	25.0265561898242\\
30	29.06016432788153\\
35	30.288151698468872\\
};
\addlegendentry{OFDM}

\end{axis}
\end{tikzpicture}%